\documentclass[aps,twocolumn,showpacs,amssymb,10pt,prb]{revtex4-1}
\usepackage{amsmath}

\usepackage{graphicx}% Include figure files
\usepackage[caption=false]{subfig}

\usepackage{bbm}
\usepackage{color}

\newcommand{\rt}{\textrm}

\newcommand{\rmd}{\rt{ d}}
\newcommand{\Id}{\rt{ Id}}
\newcommand{\HF}{\textrm{ HF}}

\newcommand{\C}{\hat{C}^{\phantom{\dagger}}}
\newcommand{\Cd}{\hat{C}^{\dagger}}
\newcommand{\vac}{|\rt{vac}\rangle}
\newcommand{\Psinot}{|\Psi_0 \rangle}
\newcommand{\LPsinot}{\langle \Psi_0 |}

\newcommand{\oa}{\hat{a}}
\newcommand{\sigcr}[2]{{[\overrightarrow{#1,#2}}]_{\Sigma}^+}  
\newcommand{\sigal}[2]{{[\overleftarrow{#1,#2}}]_{\Sigma}^-} 
\newcommand{\sigcl}[2]{{[\overleftarrow{#1,#2}}]_{\Sigma}^+}  
\newcommand{\sigar}[2]{{[\overrightarrow{#1,#2}}]_{\Sigma}^-} 
\newcommand{\sigcu}[2]{{[\overrightarrow{#1}\overleftarrow{#2}}]_{\Sigma}^+} 
\newcommand{\sigau}[2]{{[\overrightarrow{#1}\overleftarrow{#2}}]_{\Sigma}^-} 
 
\newcommand{\lbarp}[2]{\bar{\lambda}_{#1}^{#2}}
\newcommand{\pd}{\phantom{\dagger}}

\begin{document}
\title{Gutzwiller variational wave function for multi-orbital Hubbard models\\
in finite dimensions}

\author{Kevin zu M\"unster}
\author{J\"org B\"unemann}
\affiliation{Fachbereich Physik, Philipps Universit\"at Marburg,
D-35032 Marburg, Germany}

%\date{\today}

%
\begin{abstract}%
We develop a diagrammatic method for the evaluation of 
general multi-band Gutzwiller wave functions in finite dimensions.
 Our approach provides a systematic improvement of the 
 widely used Gutzwiller approximation.  As a first 
 application we investigate itinerant ferromagnetism 
 and correlation-induced deformations of the Fermi surface 
 for a two-band Hubbard model on a square lattice. %{\bf \today}
\end{abstract}

% 71.10.-w = Theories and models of many-electron system
% 71.10.Fd = Lattice fermion models (Hubbard model, etc.)
% 75.10.Lp = Band and itinerant models
% 71.18.+y = Fermi surface: calculations and measurements; effective mass, g factor
% 71.20.Be = Transition metals and alloys

\maketitle

\section{Introduction}
\label{sec:I}

Strongly-correlated electron systems display a variety of intriguing phases, such as 
superconductivity, (anti)\-ferromagnetism, or Mott insulating phases.
In order to study the fundamental properties of strongly correlated lattice systems, 
simplifying Hubbard-type models are often employed.
Unfortunately, the calculation of ground-state and dynamical properties
is notoriously difficult even for these relatively simple models.

In one dimension,
the density-matrix renormalization group (DMRG) method
permits the numerical investigation of Hubbard-type models
for fairly large chains. However, 
even modern variants of the DMRG such as 
tensor network approaches,~\cite{schollwock2011density}
are not satisfactory when applied to many-orbital models or
higher-dimensional systems.  
In the limit of infinite dimensions, 
the dynamical mean-field theory (DMFT)~\cite{RevModPhys.78.865}
maps the problem onto a single-impurity model
whose spectral function must be calculated numerically.
For multi-band systems, the solution of this task requires 
sophisticated quantum Monte-Carlo techniques and substantial 
computational resources.
Concomitantly, it is very difficult to go beyond the mean-field limit
and access multi-band models in finite dimensions.

In the absence of exact analytical or quasi-exact numerical methods,
variational approaches have proven helpful.
In this work, we employ the Gutzwiller wave function approach.\cite{Gutzwiller1964}
The evaluation of expectation values 
with the Gutzwiller correlated many-particle wave function 
poses itself a difficult many-body problem.
Therefore, even the Gutzwiller-correlated single-band Fermi sea 
can be evaluated exactly only in one 
dimension.~\cite{Metzner1987,Gebhard1987,Metzner1988,Gebhard1988,Metzner1989a,Kollar2001}
In the limit of infinite dimensions, the so-called Gutzwiller Approximation (GA) becomes 
exact for the single-band Hubbard 
model.~\cite{Metzner1987,Gebhard1990}
Later, the method was extended to the multi-band 
case.~\cite{buenemann1997b,Buenemann1997,buenemann1998,buenemann2005}
Recently, it has been combined with the density functional theory 
in a self-consistent manner to describe
transition metals and their 
compounds.~\cite{ho2008,deng2008,deng2009,wang2008,wang2010,
weng2011,PhysRevB.84.245112,PhysRevB.84.205124,PhysRevLett.111.196801,
PhysRevB.89.165122,Schickling} 

Despite many successes of the GA in 
 improving our understanding of correlated metals, there are 
 certain phenomena which it cannot describe properly. 
   For example, in single-band models the 
 Fermi surface is independent of the 
 local Coulomb interaction within the GA, unless a state with 
 broken spin or translational symmetry is considered. 
 This is obviously incorrect, as can be seen already 
 from straightforward perturbation theory for the 
 paramagnetic ground state.~\cite{schweitzer1991,halboth1997} 
 In order to describe a Fermi-surface deformation 
 one needs to evaluate the Gutzwiller wave function 
 in finite dimensions. A well established way to do this, 
 is the `variational Monte Carlo method' in which the 
 Gutzwiller energy functional is minimized numerically 
  on finite lattices.~\cite{Tocchio2008,Tocchio2011,Tocchio2012} 
Although numerically less demanding than 
 other techniques, such as  quantum Monte-Carlo,
 this method still has significant finite-size limitations. 

 An alternative approach, which has first been proposed in 
 Refs.~[\onlinecite{Gebhard1990},\onlinecite{Bunemann2012a}], 
constitutes a systematic improvement 
 of the GA for Gutzwiller wave functions on finite
 dimensional lattices. The method has been used successfully 
to study Fermi-surface deformations, d-wave superconductivity, and 
 quasi-particle band structures in single band Hubbard models,
 \cite{Bunemann2012a,PhysRevB.88.115127,doi:10.1080/14786435.2014.965235,PSSB:PSSB201552082}  
t-J models, \cite{1367-2630-16-7-073018} and periodic Anderson 
models.~\cite{PhysRevB.92.125135,marcin} 
 
Most transition metals and  their compounds
 cannot be described properly by single-band models. 
 For example,  in iron-pnictides, such as LaOFeAs, all
 five $d$-orbitals are partially occupied and may have 
 to be taken into account in any model study that aims 
 to describe the superconductivity or the antiferromagnetism in these 
systems.\cite{PhysRevLett.106.146402,PhysRevLett.108.036406}  
 Hence, it is clearly desirable to generalize the 
 method, developed  in Ref.~[\onlinecite{Bunemann2012a}] for the 
 single-band model, to the multi-band case. 
 It is the main purpose of this work, to formulate such 
 a generalization. As a first application, 
 we shall study ferromagnetism and Fermi-surface deformations 
in a two-orbital Hubbard model on a square lattice.

Our work is organized as follows. In Sect.~\ref{sec:II}, 
we introduce the multi-band Hubbard model
and the corresponding Gutzwiller wave function. Moreover, we
present a detailed derivation of the diagrammatic expansion
for ground-state expectation values.
In Sect.~\ref{sec:III} we discuss our results
for ferromagnetism and interaction-induced deformations of the Fermi surface
in a two-orbital Hubbard model on a square lattice.  
Finally, Sect.~\ref{sec:IV} summarizes our findings and gives a brief outlook.
 For some technical details we refer to three appendices.  
A more detailed derivation of the results presented 
in this work can be found in Ref.~[\onlinecite{zuMuenster2015}].

\section{Gutzwiller wave functions}
\label{sec:II}
In this work, we employ Gutzwiller variational functions for multi-band Hubbard models. 
 Such wave functions start from an independent-particle picture where the electrons are distributed over all lattice sites to optimize the single-particle energy. 
This statistical distribution leads to atomic configurations that are energetically unfavorable for finite Hubbard interactions.
In the Gutzwiller wave function, the weight of such configurations is reduced with the help of the Gutzwiller `correlator', a product of local operators, see below.
 Non-local (`extended') Gutzwiller correlators
for the single-band Hubbard model can be studied 
analytically,~\cite{Baeriswyl2009a}
and numerically using variational 
Monte Carlo.~\cite{Tocchio2008,Tocchio2011,Tocchio2012}
More recently, a two-dimensional bilayer Hubbard model 
was studied with the same method.~\cite{Ruger2014}

\subsection{Multi-band Hubbard model}

In this work, we investigate Gutzwiller-correlated wave functions for 
{\sl  general} multi-band Hubbard 
models. Only in the numerical applications in Sec.~\ref{sec:III}, 
 we shall be more specific by  considering  a two-orbital 
Hubbard model on a square lattice where the degenerate orbitals 
obey a $p_x$-$p_y$ symmetry.
The Hubbard Hamilton operator with purely local interactions reads
\begin{eqnarray}
 \hat{H} &=& \hat{H}_0 + \hat{U}\;, \\
   \hat{H}_0 &=&\sum_{i\neq j} \sum_{\sigma,\sigma'} t_{ij}^{\sigma \sigma'} 
\hat{c}_{i,\sigma}^{\dagger}\hat{c}_{j,\sigma'} \;, \\ \label{uuu}
  \hat{U} &=& \sum_{i,\sigma_{1},\ldots,\sigma_{4}} U_{\sigma_1\sigma_2\sigma_3\sigma_4} 
 \hat{c}_{i,\sigma_1}^{\dagger} \hat{c}_{i,\sigma_2}^{\dagger}
\hat{c}_{i,\sigma_3} \hat{c}_{i,\sigma_4} \;.
\end{eqnarray}
Here,  $\hat{c}_{j,\sigma}^{\dagger}$ and $\hat{c}_{j,\sigma'}$ are fermionic creation 
and annihilation operator, respectively.
The site index is given by $i$ and $j$ and the combined spin-orbital index by $\sigma$. 
The lattice indices run over all lattice sites of the lattice $\Lambda$. 
Periodic boundary conditions apply.
The hopping amplitudes $t_{ij}^{\sigma \sigma'}$ and 
the coefficients $U_{\sigma_1\sigma_2\sigma_3\sigma_4} $ of the on-site interaction 
energy are considered to be free model parameters. 

The hopping and Coulomb parameters are restricted by 
symmetry. Spin conservation and rotational symmetry  of the
$p_x$-$p_y$ orbitals reduce the nearest-neighbor and next-nearest-neighbor 
hopping amplitudes to four independent parameters.
Furthermore, the coefficients of the on-site energy 
can be expressed solely in terms of the Hubbard interaction~$U$ 
and the Hund's-rule coupling~$J$, 
as shown in appendix~\ref{appendix:lattice_symmetries}.
Note that the symmetry of the $p_x$-$p_y$ orbitals
is the same as that of the pair of $d_{xz}$-$d_{yz}$ orbitals.
Therefore, our two-band Hubbard model applies to  $p_x$-$p_y$ 
orbitals and to $d_{xz}$-$d_{yz}$ orbitals equally well.

\subsection{Definition of Gutzwiller variational states}
\label{section:definition}

The Gutzwiller correlator is given by the 
product of the local Gutzwiller correlators for all sites $l$ on our lattice~$\Lambda$, 
\begin{eqnarray}
\label{eq:mathematical_definitions:gutzwiller_global}
 \hat{P}_G = \prod_{l \in \Lambda} \hat{P}_{l} \;.
\end{eqnarray}
% A local Gutzwiller operator acts on the site $l$ only. 
If the context does not lead to any ambiguities, the local index $l$ will frequently 
 be dropped in the following. 
In this work, we restrict ourselves to the homogeneous case where the 
variational parameters in $\hat{P}_l$ are the same for all lattice sites. 
The local Gutzwiller operator is given by
\begin{eqnarray}
\label{eq:mathematical_definitions:gutzwiller_local}
 \hat{P}_{l} &=& \sum_{I_1,I_2} \lambda_{I_1,I_2} \left( |I_1 \rangle \langle 
I_2| \right)_l\;, \\\label{iio}
 \hat{P}_{l}^{\dagger} \hat{P}^{\phantom{\dagger}}_{l} 
&=& \sum_{I_1,I_2} \bar{\lambda}_{I_1,I_2}\left(  |I_1 \rangle \langle 
I_2| \right)_l\;,
\end{eqnarray}
with 
\begin{equation}
 \bar{\lambda}_{I_1,I_2}= \sum_{J} \lambda_{I_2,J}\lambda_{I_1,J} \;,
 \end{equation}
where we already assumed that the parameters $\lambda_{I_1,I_2}$ are real. 
The operators in~(\ref{eq:mathematical_definitions:gutzwiller_local}), (\ref{iio}) that act
 on the site $i$ can be written explicitly as
\begin{eqnarray}
 \left( | I_1 \rangle \langle I_2 | \right)_i = \prod_{l \in \Lambda \setminus i} \Id_l \otimes \left( | I_1 \rangle \langle I_2 | \right)_i \;,
\end{eqnarray}
where $\Id_l$ represents the identity operator on site $l$.
In our two-band application the local indices $I_1,I_2$ run over all $16$ local configurations which can contain up to four electrons.

In order to simplify the notation we define a product of local creation or annihilation operators by the introduction of the following symbols
\begin{align}\label{ssdf}
\Cd_{I} = \prod_{\sigma \in I} \hat{c}^{\dagger}_{\sigma} =  
\hat{c}^{\dagger}_{\sigma_1} \ldots \hat{c}^{\dagger}_{\sigma_n} \quad 
i<j \rightarrow \sigma_i<\sigma_j \;,\\ 
\C_{I} = \prod_{\sigma \in I} \hat{c}_{\sigma} =  
\hat{c}_{\sigma_1} \ldots \hat{c}_{\sigma_n} \quad 
i<j \rightarrow \sigma_i>\sigma_j  \;.
\end{align}
where we introduced some arbitrary order of the spin-orbit indices $\sigma$. 
The multi-particle states 
\begin{eqnarray}
 | I \rangle =\Cd_{I}| {\rm vac} \rangle 
\end{eqnarray}
are uniquely determined by the lexicographical order of their sub-indices $\sigma_i$
  in~(\ref{ssdf}).

A single particle product state (SPPS) can always be cast in the form
\begin{eqnarray}
 \Psinot = \prod_{k,\gamma}\hat{h}_{k,\gamma}^{\dagger} \vac
\end{eqnarray}
in some fermionic basis 
\begin{eqnarray}
 \hat{h}_{k,\gamma}^{\dagger} = \sum_{k,\gamma} U^{i,k}_{\sigma,\gamma}\; \hat{c}_{i,\sigma}^{\dagger} \;.
\end{eqnarray}
We will assume that the SPPS are normalized, $\langle \Psi_0  \Psinot = 1$, 
and that the canonical commutation relations hold,
$\lbrace \hat{h}_{k,\gamma}^{\dagger} ,\hat{h}_{k',\gamma'} \rbrace = \delta_{kk'} \delta_{\gamma \gamma'}$.
Now, we define the Gutzwiller wave function as
\begin{eqnarray}
 |\Psi_G \rangle = \hat{P}_G \Psinot \;.
\end{eqnarray}
In the remaining part of this work, we optimize the Gutzwiller correlator $\hat{P}_G$ and the SPPS $\Psinot$ so that the approximate ground state energy
\begin{eqnarray}\label{eng}
 E_G = \langle \hat{H} \rangle_G = \frac{\langle \Psi_G |  
\hat{H} | \Psi_G \rangle}{\langle \Psi_G |\Psi_G \rangle} 
\end{eqnarray}
becomes minimal.

\subsection{Diagrammatic expansion in finite dimensions}
We consider the expectation value of some local operator $\hat{O}_{i}$
\begin{eqnarray}
\label{eq:lct:define_expectation_value}
 \langle \hat{O}_{i} \rangle_G = \frac{\langle \Psi_G |  
\hat{O}_{i} | \Psi_G \rangle}{\langle \Psi_G |\Psi_G \rangle} \;.
\end{eqnarray}
Note that the following calculation can equally be performed for expectation values of 
 nonlocal operators such as $\langle \hat{c}^{\dagger}_{i,\sigma} \hat{c}^{\phantom{\dagger}}_{j,\sigma'}\rangle_G$.

As a first step, we follow the analysis for the single-band case derived in 
[\onlinecite{Gebhard1990},\onlinecite{Bunemann2012a}] and partly worked out for the multi-band case in infinite dimensions in [\onlinecite{Bunemann2009}]. 
In the numerator of Eq.~(\ref{eq:lct:define_expectation_value}) we pull the Gutzwiller correlators with indices $l \neq i$ to the right side of $\hat{O}_{i} $ and denote the sandwich $\hat{P}^{\dagger}_{i} \hat{O}^{\phantom{\dagger}}_{i} \hat{P}^{\phantom{\dagger}}_{i}$ as $\hat{Q}_i$,
\begin{eqnarray}
\label{eq:definition:q_operator}
 \langle \Psi_G | \hat{O}_{i} | \Psi_G \rangle 
= \langle \Psi_0 | \hat{Q}_{i} 
\prod_{l \in \Lambda\setminus i } \hat{P}^{\dagger}_{l} 
\hat{P}^{\phantom{\dagger}}_{l} | \Psi_0 \rangle\;.
\end{eqnarray}
The operator $\hat{Q}_i$ and the squares of the Gutzwiller operator $\hat{P}_l^{\dagger}\hat{P}_l$ can be written in terms of creation and annihilation operators.
 \begin{eqnarray}
 \label{eq:nodes_intermedeate}
    \hat{Q}_i &=& \sum_{ I_1, I_2 } Q'_{I_1,I_2} 
 \Cd_{i,I_1} \C_{i,I_2} \;, \\  \label{ggh}
  \hat{P}^{\dagger}_i\hat{P}_i^{\phantom{\dagger}} &=& \bar{\lambda}_{\emptyset,\emptyset}+\hat{A}_i' = \bar{\lambda}_{\emptyset,\emptyset} + \!\!\!\!\!\!  \sum_{ \substack{I_1, I_2 \\ 
|I_1|,|I_2| >0} } \!\!\!\! X_{I_1,I_2}'  \Cd_{i,I_1} \C_{i,I_2}  .
 \end{eqnarray}
where the scalar contribution $\bar{\lambda}_{\emptyset,\emptyset}$ to $\hat{P}^{\dagger}\hat{P}$ could always be
 set to unity after rescaling the Gutzwiller wave function.
However, we will postpone this step to a later stage of our analysis. In~(\ref{ggh}) 
 we have introduced the number of electrons $|I|$ in a configuration state $|I \rangle$. 

\begin{figure}[ht]
 \centering
 \includegraphics[width=0.5\textwidth]{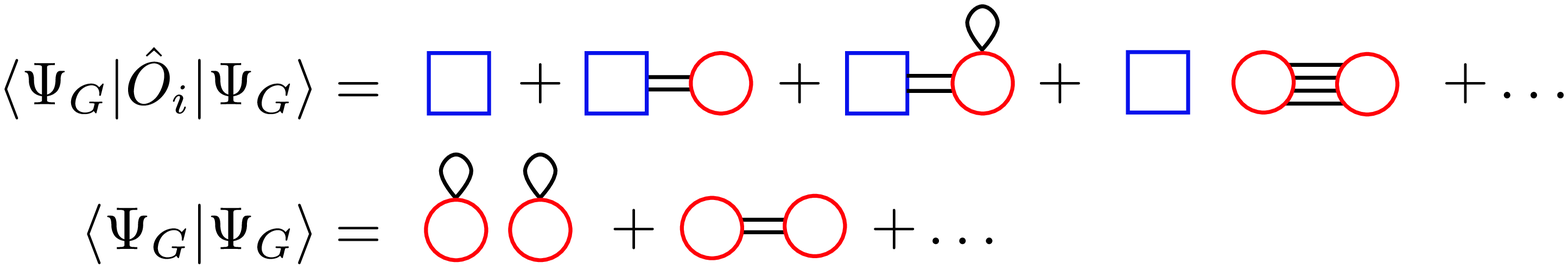}
 \caption{Diagrammatic representation of the numerator and denominator of the expectation value of an local operator $\hat{O}_i$ on the lattice site $i$. The blue square and the red circles gives the external and internal nodes, respectively. Black lines correspond to the single-particle density matrix.}
  \label{fig:introduction_chapter_two:step11}
\end{figure}

As a next step, we apply Wick's theorem 
\begin{eqnarray}
\label{eq:lct:apply_wick}
  \langle \Psi_0 | \hat{Q}_{i} 
\prod_{l \in \Lambda\setminus i } \hat{P}^{\dagger}_{l} 
\hat{P}^{\phantom{\dagger}}_{l} | \Psi_0 \rangle = \lbrace \hat{Q}_i \prod_{l \in 
\Lambda \setminus i } \hat{P}^{\dagger}_{l} 
\hat{P}^{\phantom{\dagger}}_{l} \rbrace_{\rho} \;,
\end{eqnarray}
where  $\lbrace \ldots \rbrace_{\rho}$ gives the sum over all possible contractions with respect to the density matrix 
$\rho$ with the elements
\begin{eqnarray}\label{rho}
\rho_{(i\sigma),(j\sigma')}= \rho_{ij}^{\sigma \sigma'} 
= \langle \Psi_0 | \hat{c}^{\dagger}_{i \sigma} \hat{c}^{\phantom{\dagger}}_{j \sigma'} \Psinot\;.
\end{eqnarray}
For example,
\begin{eqnarray}
  \langle \Psi_0 | \hat{c}^{\dagger}_{l\sigma_1} \hat{c}^{\pd}_{l \sigma_2} \hat{c}^{\dagger}_{k\sigma_3} \hat{c}^{\pd}_{k\sigma_4} \Psinot 
 &=&\lbrace \hat{c}^{\dagger}_{l\sigma_1} \hat{c}^{\pd}_{l\sigma_2} \hat{c}^{\dagger}_{k\sigma_3} \hat{c}^{\pd}_{k\sigma_4} \rbrace_{\rho} \\[3pt] \nonumber
 &=& \rho_{ll}^{\sigma_1\sigma_2} \rho_{kk}^{\sigma_3\sigma_4} - \rho_{lk}^{\sigma_1\sigma_4} \rho_{kl}^{\sigma_3\sigma_2}\;.
\end{eqnarray}
We depict the different contributions in a diagrammatic way, as shown in Fig.~\ref{fig:introduction_chapter_two:step11}.
Each summand of the operator $\hat{Q}$ and $\hat{P}^{\dagger}\hat{P}$ defines an `external node' 
with weight $Q'_{I_1,I_2}$ or an `internal node' with weights $X'_{I_1,I_2}$, respectively. 
Each operator contraction can be represented by a line which is either a `self-closing line'
 (also denoted as `local contractions' or `Hartree bubbles') or connects two different nodes.
In the following, we will simplify this diagrammatic analysis in three steps.

First, we aim to eliminate all local contractions.
Therefore, we map our operators to so called HF-operators, \cite{Gebhard1990,Bunemann2009} which do not have local contractions by definition.
For example,
\begin{align}
 &\langle \Psi_0 | (\hat{c}^{\dagger}_{l,\sigma_1} \hat{c}^{\pd}_{l, \sigma_2})^{HF} (\hat{c}^{\dagger}_{k,\sigma_3} \hat{c}^{\pd}_{k,\sigma_4})^{HF} \Psinot  \\ \nonumber
 &\phantom{=} = \lbrace (\hat{c}^{\dagger}_{l,\sigma_1} \hat{c}^{\pd}_{l, \sigma_2})^{HF} (\hat{c}^{\dagger}_{k,\sigma_3} \hat{c}^{\pd}_{k,\sigma_4})^{HF}  \rbrace_{\rho}\\ \nonumber
  &\phantom{=} =  - \rho_{lk}^{\sigma_1\sigma_4} \rho_{kl}^{\sigma_3\sigma_2} \;.
\end{align}
The mapping between the normal creation and annihilation operators and the HF-operators depends on the local density matrix $\rho_{ij}^{\sigma \sigma'}$ as can be seen from the simplest case
\begin{eqnarray*}
( \hat{c}^{\dagger}_{i,\sigma} \hat{c}^{\pd}_{i,\sigma'} )^{\rt{HF}} &= \hat{c}^{\dagger}_{i,\sigma} \hat{c}^{\pd}_{i,\sigma'} - \rho_{ii}^{\sigma\sigma'}\;.
\end{eqnarray*}
An extension of this mapping for operator products $\Cd_{i,I_1} \C_{i,I_2}$ is given in appendix~\ref{section:appendix:hf_operators}.
We write the operator $\hat{Q}$ and the square of the Gutzwiller correlator as
 \begin{eqnarray}
    \hat{Q}_i &=& \sum_{ I_1, I_2 } Q_{I_1,I_2} 
\left( \Cd_{i,I_1} \C_{i,I_2} \right)^{  \rt{HF} }\;, \\
  \hat{P}^{\dagger}_i\hat{P}^{\phantom{\dagger}}_i &=& 1+\hat{A}_i = 1 + \!\!\!\!\!\!  \sum_{ \substack{I_1, I_2 \\ 
|I_1|,|I_2| >0} } \!\!\!\! X_{I_1,I_2} \left( \Cd_{i,I_1} \C_{i,I_2} 
\right)^{ \rt{HF}} \!\!,
 \end{eqnarray}
where we set the coefficient $X_{\emptyset,\emptyset}=1$. 
As mentioned above, this is equal to a rescaling of the Gutzwiller wave function by a factor 
which is always canceled out by the denominator in Eq.~(\ref{eq:lct:define_expectation_value}).

\begin{figure}[ht]
 \centering
 \includegraphics[scale=0.45]{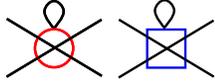}
 \caption[All nodes with internal lines cancel out.]
 {All nodes with internal lines cancel because the local entries of $\bar{\rho}_{ii}^{\sigma \sigma'}$ are set to zero.}
  \label{fig:introduction_chapter_two:step22}
\end{figure}

All operators in Eq.~(\ref{eq:lct:apply_wick}) are normal ordered because all site 
indices are different when we apply Wick's theorem.
We can set all local entries in $\rho_{ii}^{\sigma \sigma'}$ to zero because we work 
with the HF-operators so that 
all local contractions vanish automatically.
Therefore, we can carry out all contractions with a new density matrix 
\begin{eqnarray}
 \bar{\rho}^{\sigma \sigma'}_{ij} =  \rho^{\sigma \sigma'}_{ij} - \delta_{ij} \;  \rho^{\sigma \sigma'}_{ii}
\end{eqnarray}
and drop the HF-operator notation at the same time
\begin{eqnarray}
  \lbrace \left( \hat{C}_{i,I_1}^{\dagger} \C_{i,I_2} \right)^{ \rt{ HF} }  
\ldots \rbrace_{\rho} \equiv 
  \lbrace  \hat{C}_{i,I_1}^{\dagger} \hat{C}^{\phantom{\dagger}}_{i,I_2}    \ldots 
\rbrace_{\bar{\rho}} \;.
\end{eqnarray}
Without any nonzero local contraction we get
\begin{eqnarray}
\label{eq:lct:grassmann}
  \lbrace \hat{c}^{\dagger}_{i,\sigma} \hat{c}^{\phantom{\dagger}}_{i,\sigma} \ldots 
\rbrace_{\bar{\rho}} = - \lbrace  \hat{c}^{\phantom{\dagger}}_{i,\sigma} \hat{c}^{\dagger}_{i,\sigma} 
\ldots \rbrace_{\bar{\rho}} \;.
\end{eqnarray}
 Thus, we can replace the fermionic operators $\hat{c}_{i,\sigma}$, $\hat{C}_{i,I}$ 
 by Gra\ss mann variables $\tilde{c}_{i,\sigma}$, $\tilde{C}_{i,I}$, respectively. 
 These Gra\ss mann variables are nilpotent
 \begin{eqnarray}
  \tilde{C}_{i,I_1} \tilde{C}_{i,I_2} = 0 \quad \rt{ if } I_1 \cap I_2 \neq 
\varnothing\;.
\end{eqnarray}
In principle, the introduction of the HF mapping is not a necessary step for the introduction of Gra\ss mann operators as we discuss in appendix~\ref{section:appendix:previous_approaches}.
All local entries $\bar{\rho}_{ii}^{\sigma \sigma'}$ of the new density matrix vanish so that the diagrammatic expansion cannot have nodes with self-closing lines, as shown in Fig.~\ref{fig:introduction_chapter_two:step22}.

The coefficients $Q_{I_1,I_2}$ and $X_{I_1,I_2}$  are not affected by our mapping so that we can write
\begin{eqnarray}
\tilde{Q}_i &=& \sum_{ I_1, I_2  } Q_{I_1,I_2}  
\tilde{C}^{\dagger}_{i,I_1} \tilde{C}^{\phantom{\dagger}}_{i,I2} \\
\tilde{A}_i &=& \!\!\!\!\!\! \sum_{ \substack{I_1, I_2 \\ 
|I_1|,|I_2| >0} } \!\!\!\!  X_{I_1,I_2}  \tilde{C}^{\dagger}_{i,I_1} \tilde{C}^{\phantom{\dagger}}_{i,I2} \;.
\end{eqnarray}
The numerator in Eq.~(\ref{eq:lct:apply_wick}) becomes
\begin{eqnarray}
\label{eq:lct:numerator_with_grassmann_variables}
 \lbrace{ \hat{Q}_i\prod_{l \in \Lambda\setminus i }  (1+\hat{A}_l) \rbrace} 
{}_{\rho} 
  = \lbrace{  \tilde{Q}_i\prod_{l \in \Lambda\setminus i  }  (1+\tilde{A}_l) 
\rbrace} {}_{\bar{\rho}} 
\end{eqnarray}
whereas the denominator reads
\begin{eqnarray}
\label{eq:lct:denominator_with_grassmann_variables}
 \lbrace{ \prod_{l \in \Lambda }  (1+\hat{A}_l) \rbrace} 
{}_{\rho} 
  = \lbrace{  \prod_{l \in \Lambda  }  (1+\tilde{A}_l) 
\rbrace} {}_{\bar{\rho}} \;.
\end{eqnarray}

As a second step, we merge the diagrams of the numerator and denominator with the help of the linked cluster theorem.
To this end, the lattice site restrictions on the right hand site of Eq.~(\ref{eq:lct:numerator_with_grassmann_variables}) must be removed. 
Therefore, we define 
\begin{eqnarray}
\label{eq:lct:exponential_form_internal}
  1+\tilde{A}_i = \exp(\tilde{G}_i) \;, \\
  \label{eq:lct:exponential_form_internal2}
   \tilde{Q}_i = \tilde{M}_i \exp(\tilde{G}_i)   \;.
\end{eqnarray}
with
\begin{eqnarray}
  \tilde{G}_i &=& \sum_{ \substack{I_1,I_2 \\ |I_1|,|I_2| >0}} Z_{I_1,I_2} \tilde{C}^{\dagger}_{i,I_1} \tilde{C}^{\phantom{\dagger}}_{i,I_2} \;, \\
    \tilde{M}_i &=& \sum_{I_1,I_2 } M_{I_1,I_2} \tilde{C}^{\dagger}_{i,I_1} \tilde{C}^{\phantom{\dagger}}_{i,I_2} \;.
\end{eqnarray}
The exponential series expansion stays finite due to the nilpotency of the Gra\ss mann variables. 
Therefore,  the new coefficients $Z_{I_1,I_2}$ and $M_{I_1,I_2}$ can be written as finite polynomials of the old coefficients $Q_{I_1,I_2}$ and $X_{I_1,I_2}$. The explicit expressions are given in appendix~\ref{section:appendix:hf_operators}.
It is crucial that we perform the HF-mapping before we switch to the exponential form of our correlators. 

These additional redefinitions allow us to cast Eq.~(\ref{eq:lct:numerator_with_grassmann_variables}) into the form
   \begin{eqnarray}
   \label{eq:lct:numerator_in_exponential_form}
     \langle \Psi_G| \hat{O}_i | \Psi_G \rangle   = \lbrace  \tilde{M}_i \prod_{ l \in \Lambda } \exp(\tilde{G_l}) 
\rbrace_{\bar{\rho}} \;.
  \end{eqnarray}
Note that the site index restriction $l \neq i$ disappeared. 
Eq.~(\ref{eq:lct:denominator_with_grassmann_variables}) can be rewritten as
   \begin{eqnarray}
   \label{eq:lct:denominator_in_exponential_form}
     \langle \Psi_G| \Psi_G \rangle  = 
    \lbrace  \prod_{ l \in \Lambda } \exp(\tilde{G_l}) 
\rbrace_{\bar{\rho}} \;.
  \end{eqnarray}

\begin{figure}[ht]
 \centering
 \includegraphics[width=0.42\textwidth]{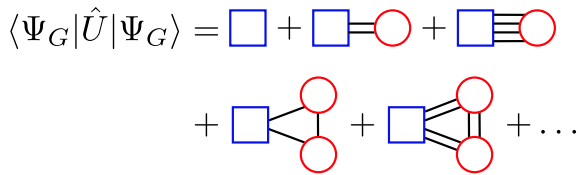}
 \caption[The first few connected diagrams that contribute to $\langle \hat{U}\rangle_G$.]
 {The first few connected diagrams that contribute to $\langle \hat{U}\rangle_G$. 
 The blue square represents the external node. The red circles represent the internal nodes.
 Black lines stand for the single-particle density matrix.
  The second and the forth diagram cancel out after the introduction of a 
gauge in the variational parameters $\lambda_{I_1,I_2}$ as shown in 
subsection \ref{subsection:infinite_d}.}
 \label{fig:lct:diagrammatic_expansion}
\end{figure}

Now, we are in the position to apply the linked cluster theorem (LCT), as described, e.g., 
in~[\onlinecite{fetter2003quantum}]. 
We find
\begin{eqnarray}
\label{eq:lct:lct}
 \frac{ \langle \Psi_G|  \hat{O}_i | \Psi_G \rangle }{ \langle \Psi_G| \Psi_G 
\rangle } =  \sum_{L\subset \Lambda }\frac{1}{|L|!} \; \lbrace  
\tilde{M}_i \prod_{l\in L} \tilde{G_{l}} \rbrace_{\bar{\rho}}^{\rt{conn.}} \;,
\end{eqnarray}
where the summation is performed over all subsets $L$ of the lattice $\Lambda$. 
The first few diagrams that are needed for the evaluation of the potential energy are shown in Fig.~\ref{fig:lct:diagrammatic_expansion}.

Some of the polynomials of $\tilde{G}_{l}$ in Eq.~(\ref{eq:lct:numerator_in_exponential_form}) vanish due to the nil-potency of the Gra\ss mann variables. 
In contrast to the usual application of the LCT in many-body lattice theories, we can apply our expansion for a finite lattice as well. 
The nil-potency property allows us to add virtually as many nodes as we need to regroup our diagrams in all orders.
Note that after the application of the LCT the nodes $\tilde{G}_{l}$ are contracted in such a way that all nodes have to be connected to the external nodes $M_i$. 
This invalidates the nil-potency of the Gra\ss mann variables inside the curly brackets.
Therefore, several nodes can be located on the same site as long as these nodes are only connected indirectly.
In a third step we will eliminate all internal nodes with two lines as described in the next section.

\subsection{Limit of infinite dimensions}
\label{subsection:infinite_d}
\begin{figure}[ht]
 \centering
 \includegraphics[scale=0.45]{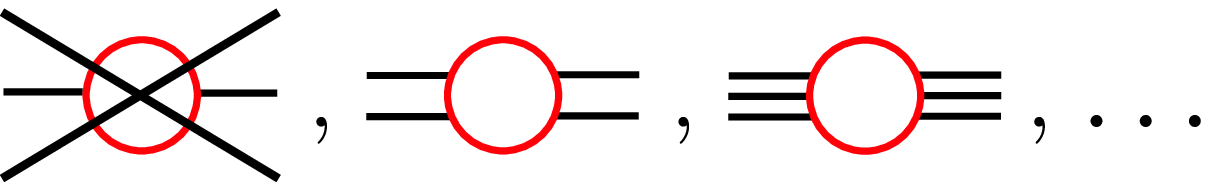}
 \caption[Elimination of two-line nodes.]
 {The internal nodes with only two lines will be eliminated by a gauge in the variational parameters.}
  \label{fig:introduction_chapter_two:step32}
\end{figure}

A scaling analysis of the `kinetic energy operator' $\hat{H}_0$ 
shows\cite{Metzner1989a,Gebhard1990,Gebhard1990} that the lines of the density matrix scale with the lattice dimension $d$ as
\begin{eqnarray}
\rho_{ij} \sim (\sqrt{2d})^{-||i-j||_{\rt{1}}}\;,
\end{eqnarray}
where we dropped the spin-band index for notational clarity, and $||.||_1$ gives the `one-norm' (or `Manhattan metric') of the displacement vector $i-j$.
One can show that all diagrams vanish if two nodes are connected to each other by at least three independent paths. 
This means, that there are at least three distinctive paths $\rho_{aj_1} \rho_{j_1 j_2}\ldots \rho_{j_m b}$ from node $(a)$ to node $(b)$, so that none of the subsegments $\rho_{ij}$ coincide.
A trivial example is the diagram in which two internal nodes are connected by at least three independent lines.
It will even be possible to eliminate all nontrivial diagrams if the nodes that have a single outgoing and incoming line are eliminated, as shown in Fig.~\ref{fig:introduction_chapter_two:step32}.
For this reason, a gauge in the variational parameters is introduced which sets the weight of these nodes to zero
\begin{eqnarray}
\label{eq:gw_constraint}
 Z_{I_1,I_2} = 0 \quad \forall |I_1|=|I_2|=1 \;.
\end{eqnarray}
This constraint must be incorporated in the optimization of the variational parameters $\lambda_{I_1,I_2}$.
However, we can show that this constraint will not reduce the variational freedom in our model.\cite{zuMuenster2015}

\begin{figure}[ht]
 \centering
 \includegraphics[scale=0.45]{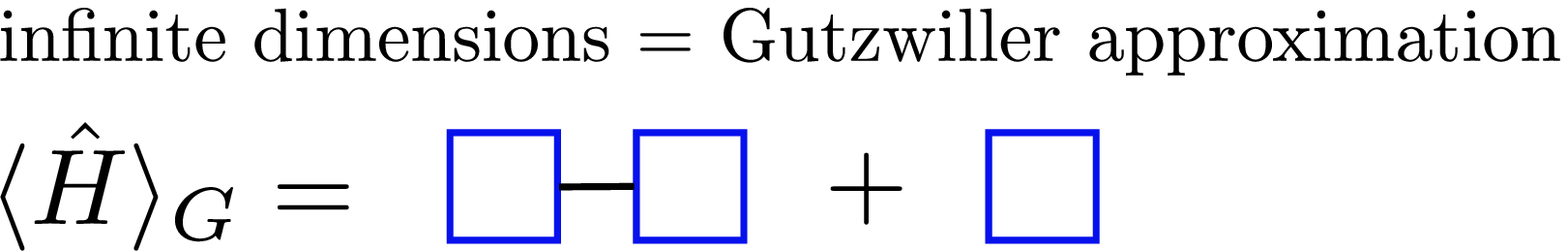}
 \caption{Diagrammatic analysis in infinite dimensions.}
 \label{fig:introduction_chapter_two:infinite_and_second_order_22}
\end{figure}

Then, the scaling of the hopping parameters 
\begin{eqnarray}
 t_{ij}  \sim (\sqrt{2d})^{-||i-j||_{\rt{1}}}
\end{eqnarray}
shows that all contributions with an internal node or two external nodes that are 
 connected by three or more lines scale at least as $\sim 1/d$. 
In the limit $d\to \infty$, the only remaining terms are given by
\begin{eqnarray}
\label{eq:lct:infinite_d_limit}
 \langle\hat{H}_0\rangle_G &=& \sum_{i\neq j} \sum_{\sigma\sigma',\tau\tau'} \!\!\!\! M_{\sigma,\emptyset}(\hat{c}_{\tau}^{\dagger}) M_{\emptyset,\sigma'}(\hat{c}_{\tau'}) \; t_{ij}^{\tau \tau'} \rho_{ij}^{\sigma \sigma'} \\ 
 \langle\hat{U}\rangle_G &=& M_{\emptyset}(\hat{U}) \;,
\end{eqnarray}
as shown in Fig.\ref{fig:introduction_chapter_two:infinite_and_second_order_22}.

 In the rest of our work we will refer to the terms in Eq.~(\ref{eq:lct:infinite_d_limit}), 
 already derived in~[\onlinecite{buenemann1998}],  as the `infinite-$d$ limit'.
From these results we conclude that the constraints~(\ref{eq:gw_constraint}) ensure that the 
 leading order terms of our diagrammatic expansion 
correspond to the exact Gutzwiller ground-state energy in infinite dimensions.
 As demonstrated for the single-band model in one dimension,\cite{Bunemann2012a}  the  constraints are 
 also essential for a rapid convergence of our expansion beyond the infinite-$d$ limit.

\subsection{Optimization of $\Psinot$}
In this work, we use the optimization algorithm which was introduced in~[\onlinecite{Bunemann2012a}]. 
 The energy~(\ref{eng}) depends on the variational parameters $\lambda_{I_1,I_2}$ and the state 
 $\Psinot $, where the latter enters the functional only through the single-particle
 density matrix~(\ref{rho}),  
\begin{eqnarray}\label{ssd}
E_{\rm G}=E_{\rm G}(\lambda_{I_1,I_2},\rho^{\sigma\sigma'}_{ij}).
 \end{eqnarray}
As shown, e.g., in appendix A of Ref.~[\onlinecite{Schickling}], the minimization of  $E_{\rm G}$ 
with respect to  $\rho$ leads to the following 
effective single-particle Hamiltonian
\begin{eqnarray}
  \label{eq:ps_optimization:effective_hamilton_operator}
  \hat{H}^{\rt{eff}}_0  &=&\sum_{i\neq j} \sum_{\sigma\sigma'} t_{ij}^{{\rm eff};\sigma \sigma'} 
  \hat{c}^{\dagger}_{i\sigma}\hat{c}_{j\sigma'} \;, \\\label{qwe}
 t_{ij}^{{\rm eff};\sigma \sigma'} &=&\partial_{\rho_{ij}^{\sigma \sigma'}} E_{\rm G} \;.
\end{eqnarray}
which has $|\Psi_0\rangle$ as a  ground state.  
Hence, for the minimization of~(\ref{ssd}) we need to solve
\begin{eqnarray}\label{1234}
  \hat{H}_{\rt{eff}} \Psinot = E_S \Psinot \;,
\end{eqnarray}
and minimize $E_{\rm G}$ with respect to $\lambda_{I_1,I_2}$,
\begin{eqnarray}\label{123}
\partial_{\lambda_{I_1,I_2}} E_{\rm G}=0\;.
\end{eqnarray}
In order to solve these equations self-consistently, we usually start with the 
ground state $\Psinot^{(0)}$ of the free system, i.e., we set 
$t_{ij}^{{\rm eff};\sigma \sigma'}=t_{ij}^{\sigma \sigma'}$ in~(\ref{qwe}),(\ref{1234}).
Then we compute the density matrix, solve~(\ref{123}), and determine new 
 parameters~(\ref{qwe}). 
The optimization terminates if the change of the  effective hopping parameters 
  between two cycles drops below some threshold.
In order to test the stability of the algorithm, we can start from a different initial state.
This initial state may be constructed from a perturbed kinetic energy operator 
$\hat{H}_0+\delta \hat{h}_0$.
Usually the optimization algorithm remains stable against these perturbations but in some cases 
the fix-point of this map does not need to be unique, as  shown 
in~[\onlinecite{Bunemann2012a}] where 
a  symmetry breaking of the Fermi surface (Pomeranchuk phase) has been investigated.

\section{Results}
\label{sec:III}

\subsection{Magnetism}
The occurrence of a ferromagnetic phase is favored by two conditions. 
The local Hamiltonian favors the formation of local magnetic moments for positive values of the Hund's-rule coupling 
$J$. 
Then, the two-particle eigenstates of the on-site energy $\hat{U}$ with maximal local spin $S=1$ are lowest in energy, 
in accordance with Hund's first rule.
Therefore, for large values of $J$, the ground state of the lattice system may show global ferromagnetism
 if the pre-formed local moments align.
In contrast to that, the Stoner picture gives a different explanation for the origin of ferromagnetism.
In this picture, a splitting between majority and minority bands reduces their mutual Coulomb repulsion due to
 the Pauli principle. 
This effect becomes significant when the density of states $D(E_{\rt{F}})$ at the Fermi energy is large.
In the Gutzwiller variational approach, the number of energetically costly multiple occupancies is reduced by an 
adjustment of the variational parameters. 
Therefore, we can expect that the Gutzwiller wave function predicts ferromagnetism at much larger interaction
 strengths than the uncorrelated SPPS. 
The Gutzwiller variational description leaves room both for the Stoner band splitting and the local moment 
formation as a source for itinerant ferromagnetism.

\begin{figure}[ht]
 \centering
\includegraphics[width=0.35\textwidth]{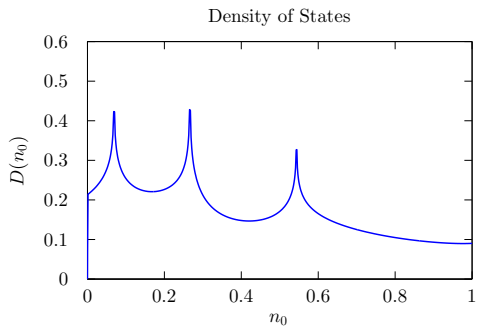}
\caption[The density of states $D(n_0)$ as a function of the density $n_0$.]
{The density of states $D(n_0)$ as a function of the density $n_0$.
The hopping amplitudes are $ t_{x}^{11} = -1.0 ,\; t_{y}^{11} = -0.6, \;
 t^{11}_{xy}= 0.2, \; t^{12}_{xy}=-0.4$.}
\label{fig:ferromagnetism:dos1}
\end{figure}

Throughout this section, we focus on the kinetic energy operator with some generic amplitudes
 \begin{eqnarray}
 t_{x}^{11} = -1.0 ,\; t_{y}^{11} = -0.6 ,\; t^{11}_{xy}= 0.2 ,\; t^{12}_{xy}=-0.4 \;,
\end{eqnarray}
where the coefficients $t_{x}^{11}$ and $t_{y}^{11}$ give the hopping amplitudes for a transition process 
between two $p_x$ orbitals to its horizontal and vertical neighbors, respectively. 
The coefficients $t^{11}_{xy}$ and $t^{12}_{xy}$ give the hopping amplitude for the transition process to 
the next-nearest neighbors with and without an inter-orbital transition. 
A detailed analysis of the lattice symmetries and all remaining hopping coefficients can be found in 
appendix~\ref{appendix:lattice_symmetries}.

The density of states $D(n_0) = D(E)|_{E=E_\rt{F}(n_0)}$ in Fig.~\ref{fig:ferromagnetism:dos1} shows 
three peaks at $n_0 \approx 0.069$, $0.266$ and $0.543$.
Below we investigate the ferromagnetic transition at $n_0=0.2$, $0.265$, $0.275$, $0.3$.
The densities are located near the second peak in the density of states.
The corresponding kinetic energies of the free system $E_{\rt{kin}}^0 = \LPsinot \hat{H}_0 \Psinot$ 
are $E_{\rt{kin}}^0 = -1.8172$, $-2.2639$, $-2.3274$ and $-2.4797$, respectively.
Therefore, the average kinetic energy is very similar for all cases under investigation. 

We define the quantity $M$ as
\begin{eqnarray}
 \langle \hat{n}_{i\uparrow} \rangle &=& n_{\uparrow} = (1+M)n_0  \;, 
\\ \nonumber
 \langle \hat{n}_{i\downarrow} \rangle &=& n_{\downarrow}=(1-M)n_0
\end{eqnarray}
with $i \in \{1,2\}$, so that $0\leq M \leq 1$, and the total density remains constant.
The total magnetization will be given by $M_{\rt{tot}} =  2(n_{\uparrow}- n_{\downarrow})$ when both bands are considered. 
In order to obtain the optimal magnetization we will perform a scan in the $M$-$U$ plane while we keep the ratio of $J$ and $U$ fixed to $J/U=8/30$. 
Then, we optimize the band symmetric Gutzwiller variational parameters $\lambda_{I_1,I_2}$ for each magnetization.
Our diagrammatic expansion includes all lines $\rho_{ij}^{\sigma\sigma'}$ with $ ||i-j||_{1} \leq 4$.

\begin{figure}[ht]
\includegraphics[scale=0.245]{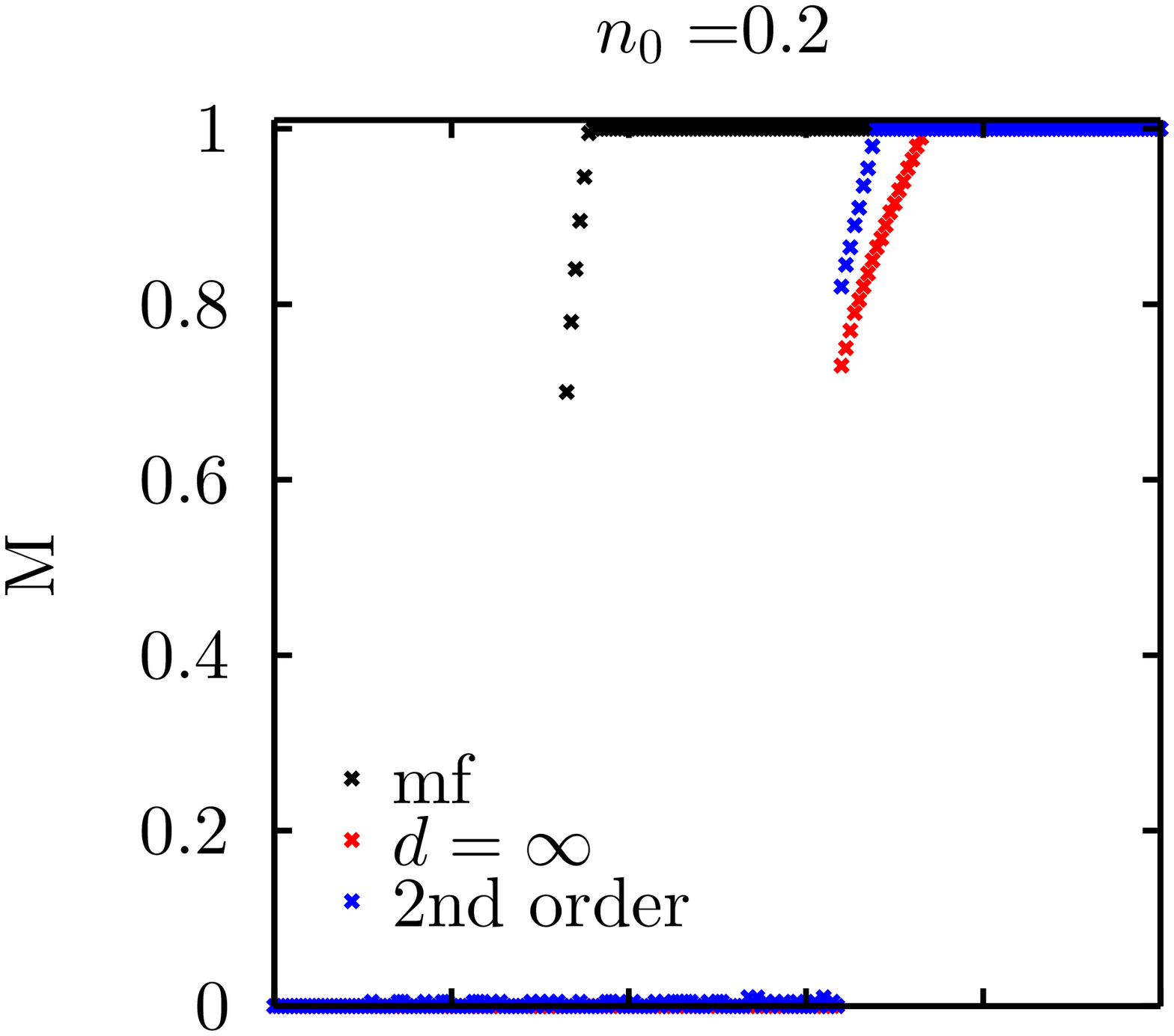} 
\includegraphics[scale=0.245]{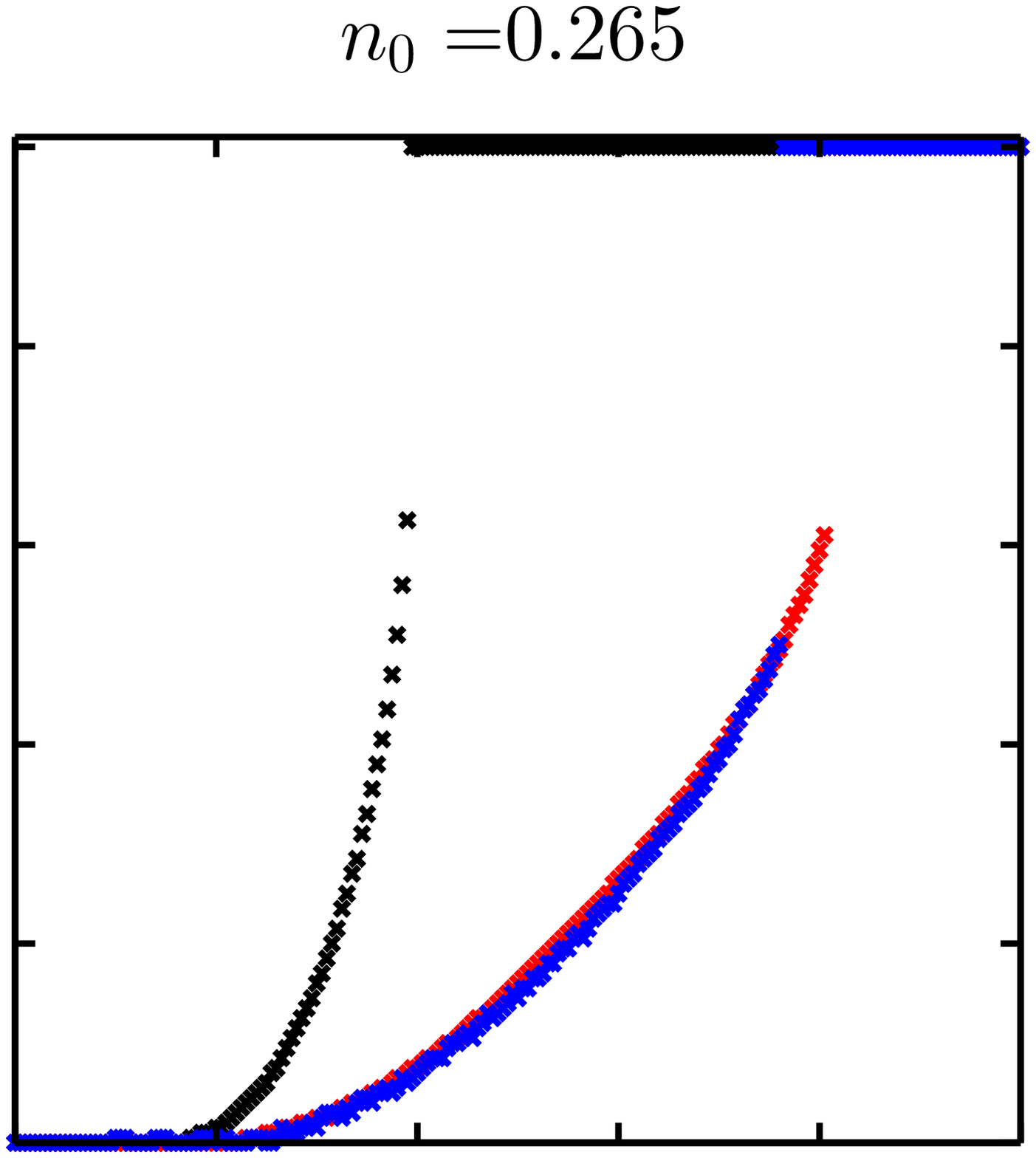} \\
\includegraphics[scale=0.245]{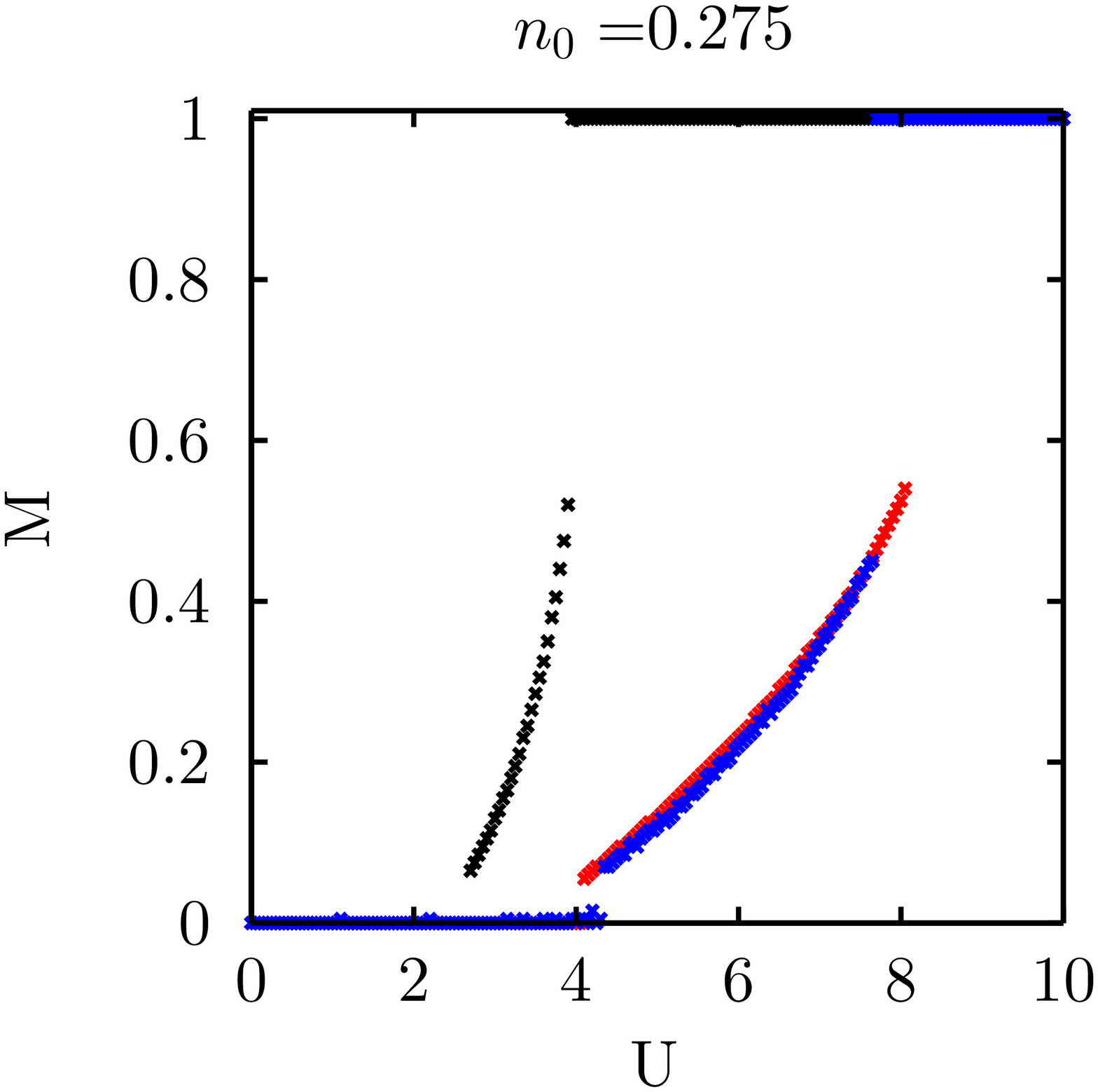} 
\includegraphics[scale=0.245]{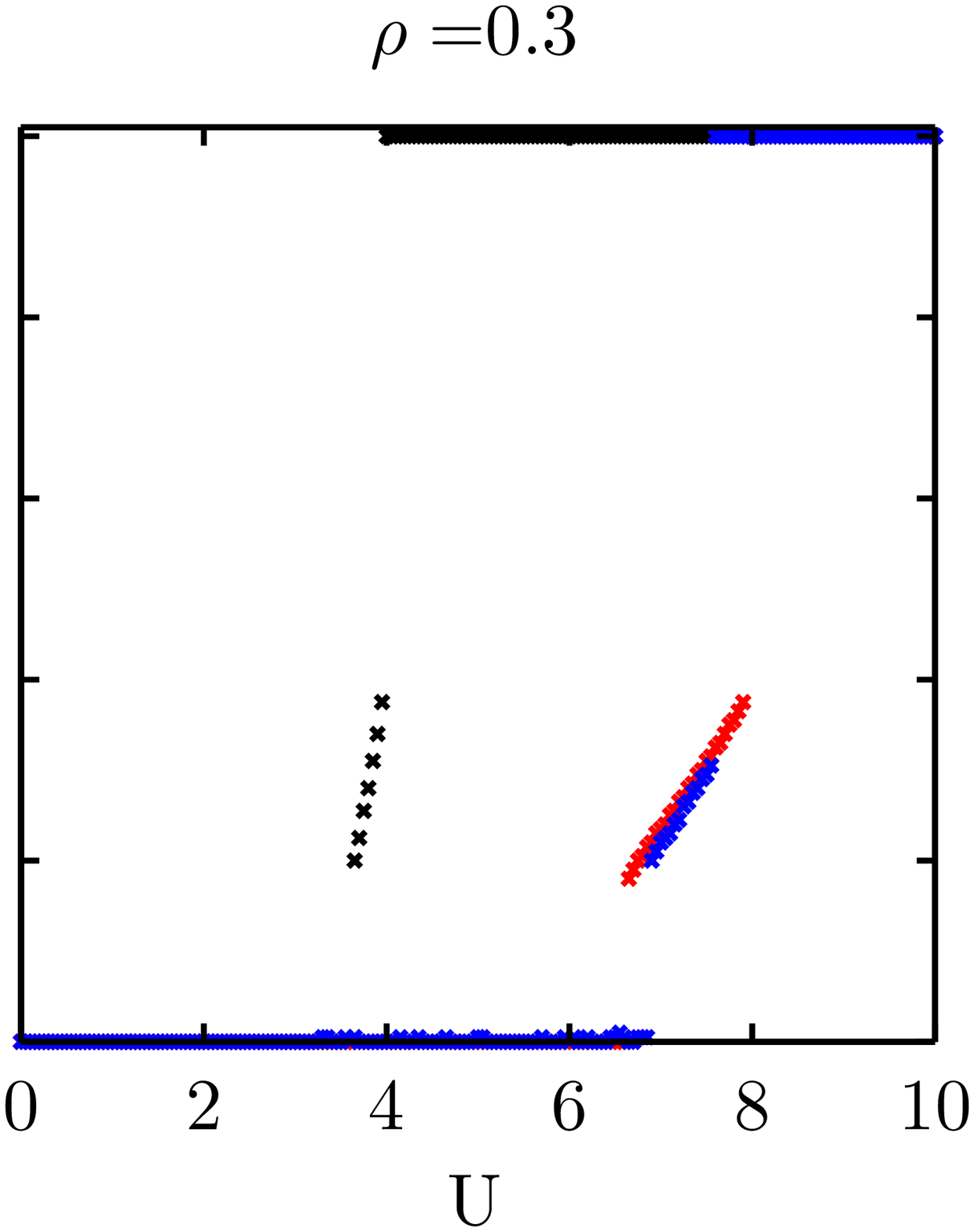}
\caption[Magnetization of the Gutzwiller wave function.]
{Magnetization of the Gutzwiller wave function. 
The black, red and blue crosses give the HF, the infinite-$d$ and the second order approximation, respectively. 
}
\label{fig:ferromagnetism:magnetization1}
\end{figure}

As seen in Fig.~\ref{fig:ferromagnetism:magnetization1}, in the Hartree-Fock approximation the magnetization jumps 
to a finite value at $U_{\rt{HF}} \approx 3.3$.
The magnetization then increases monotonically until the ground-state is fully polarized at $U_{\rt{HF}}^{\rt{sat.}}\approx 3.6$.
A detailed analysis of the ground state energy shows that the nature of the jumps can be understood as a first-order 
phase transition.
The Gutzwiller approach reveals a different picture.
In second order, the magnetization  shows a finite magnetization for $U_{\rt{G}} \approx 6.4$ and becomes fully 
magnetized for $U_{\rt{G}}^{\rt{sat.}} \approx 6.68$. 
This shows that the magnetization is shifted to much larger interaction strengths in the Gutzwiller wave 
function.
The infinite-$d$ approximation becomes magnetized at $U_{\rt{G}}^{\infty}\approx 6.4$ and becomes fully 
magnetized for $U_{\rt{G}}^{\infty,\rt{sat.}}\approx 7.35$.
Therefore, the second order diagrams in our diagrammatic expansion do not change the results on
 ferromagnetism significantly.

Next, we analyze the density $n_0=0.265$ that lies very close to the second peak in the density of states, see Fig.~\ref{fig:ferromagnetism:dos1}.
In the Stoner picture, the large density of states causes a finite magnetization at much smaller interaction strength.
In our second-order Gutzwiller approach, the ground-state becomes already magnetized at $U_{\rt{G}}\gtrsim 2$ although a precise evaluation of the 
threshold is hindered by numerical difficulties. 
The magnetization in the second-order approximation jumps to the fully magnetized state at $U_{\rt{G}}^{\rt{sat.}} = 7.6$.
The infinite-$d$ approximation lies almost on top of the second-order expansion except at the transition to the fully magnetized state which occurs at $U_{\rt{G}}^{\infty,\rt{sat.}}\approx 8.1$.
The HF-result shows the same qualitative behavior but the onset of ferromagnetism is at $U_{\rt{HF}} < U_{\rt{G}}$.
Moreover, the magnetization increases more rapidly as function of the interaction strengths and saturates already at $U_{\rt{HF}}^{\rt{sat.}} \approx 3.95$.

For $n_0=0.275$ the second-order magnetization result jumps to a finite value at $U_{\rt{G}} \approx 4.3$ and becomes fully spin polarized at  $U_{\rt{G}}^{\rt{sat.}} \approx 7.6$.
The transition points of the infinite-$d$ (HF) approximation lie at $U_{\rt{G}}^{\infty} \approx 4.1$ ($U_{\rt{HF}} \approx 2.7$) and $U_{\rt{G}}^{\infty,\rt{sat.}}\approx 8.05$ ($U_{\rt{HF}}^{\rt{sat.}} \approx 3.95$), respectively.
The magnetization curve shows the same qualitative behavior in all three approximations:
The magnetization jumps to a small but finite value. 
Then the magnetization increases gradually as a function of $U$ whereby the slope is much steeper in HF than in Gutzwiller theory.
Lastly, the magnetization jumps to full saturation at $U^{\rt{sat.}}$.
In general, the critical values are much larger in Gutzwiller theory than in the Hartree-Fock approach.
Note that the second-order terms to the result in $d=\infty$ lead to fairly small quantitative corrections.
The magnetization onset requires a larger interaction strength $U$ for $n_0=0.275$ because the density of states is lower for $n_0 = 0.275$ than for $n_0 = 0.265$.
Furthermore, we can see that the transitions to the fully magnetized state occur at almost the same interaction strength as for $n_0 = 0.265$ in all approximations.
This shows, that the transition to the fully polarized state depends on the density but not on the density of states.

For $n_0=0.3$ we still recover qualitatively the same behavior as for $n_0=0.275$ but the region between the onset of ferromagnetism and the transition to the fully polarized phase becomes smaller.
Again, the critical values in Gutzwiller theory are about a factor of two larger than in Hartree-Fock theory.

In summary we can state that a large density of states at the Fermi energy promotes ferromagnetism.
The Gutzwiller approach shows, however, that ferromagnetism, in general, requires large Coulomb interactions.
Moreover, the Gutzwiller approach leaves room in parameter space for non-saturated ferromagnetism.
 For the system parameters, considered in this work,  
phases with long-range magnetic order are already
 well described within the GA. This supports the use of this approximation in many earlier works, 
 see, e.g.,  Refs.~[\onlinecite{Schickling},\onlinecite{PhysRevLett.106.146402},\onlinecite{PhysRevLett.108.036406}].  

\subsection{Fermi surface deformations}
In this subsection we show that the optimization of the SPPS can lead to a deformed Fermi surface.
In some cases, the Fermi surface even changes its topology.
Note that Fermi surface deformations within the Gutzwiller variational approach 
  have already been studied in~[\onlinecite{Bunemann2012a}] for the single-band Hubbard model. 

For our degenerate Hubbard model in the infinite-$d$ limit, neither 
for the Gutzwiller wavefunction nor within a more sophisticated DMFT calculation, 
we would find any correlation-induced changes of the 
Fermi-surface. Hence, all results in this section can  be understood as an effect of the 
finite-dimensional evaluation provided by our higher-order expansion.

\begin{figure}[ht]
 \centering
\includegraphics[width=0.35\textwidth]{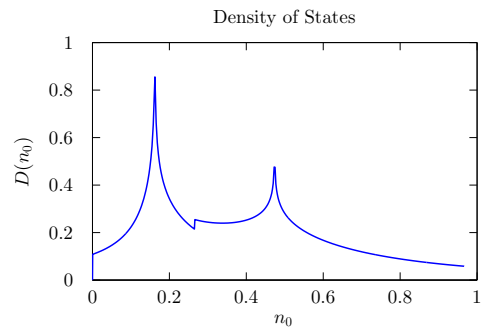}
\caption[The density of states $D(n_0)$ as a function of the density $n_0$.]
{The density of states $D(n_0)$ as a function of the density $n_0$.
The hopping amplitudes are $ t_{x}^{11} = -1.0 \;,\;\; t_{y}^{11} = -0.5 \;,\;\;
 t^{11}_{xy}= 0.4 \;,\;\; t^{12}_{xy}=-0.2$.}
\label{fig:ps_optimization:dos2}
\end{figure}

We examine the Fermi-surface deformations for the following parameter set,
 \begin{eqnarray}
 t_{x}^{11} = -1.0 ,\;t_{y}^{11} = -0.5 ,\;
 t^{11}_{xy}=0.4 ,\; \\ \nonumber 
 t^{12}_{xy}=-0.2, \; U=6.0\;,\;\; J=0.8\;.
\end{eqnarray}
The amplitudes are chosen in such a way that the topology of the Fermi surface changes near half-filling where the effect of the Gutzwiller correlator is strongest.
The density of states is shown in Fig.~\ref{fig:ps_optimization:dos2}.
The peak near $n_0 = 0.47$ is caused by the change in the topology of the Fermi surface.

\vspace{-8pt}

Our diagrammatic expansion includes all lines $\rho_{ij}^{\sigma\sigma'}$ with $ ||i-j||_{1} \leq 5$.
In some cases, the optimization algorithm alternates between two fix points which are energetically very close. 
However, the Fermi surface of these fix points may differ significantly. 
In these cases it is useful to introduce some damping for the effective hopping parameters 
 (\ref{qwe}) in the self-consistency cycle. 
In our calculations, we usually find a fix-point after $n<15$ iteration steps. 

\begin{figure}[t]
 \centering
\vspace{0pt}\includegraphics[scale=0.245]
{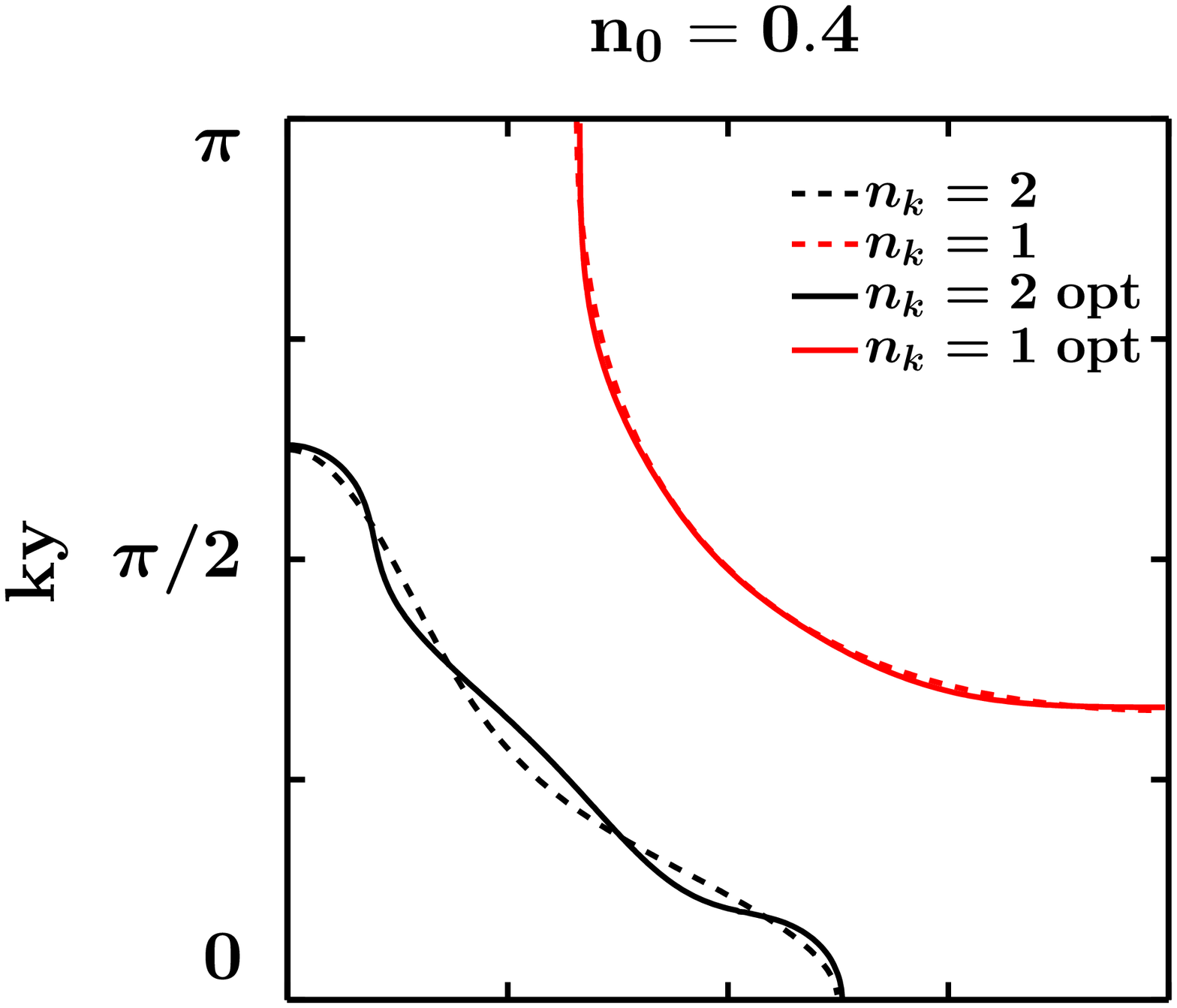}
\vspace{0pt}\includegraphics[scale=0.245]
{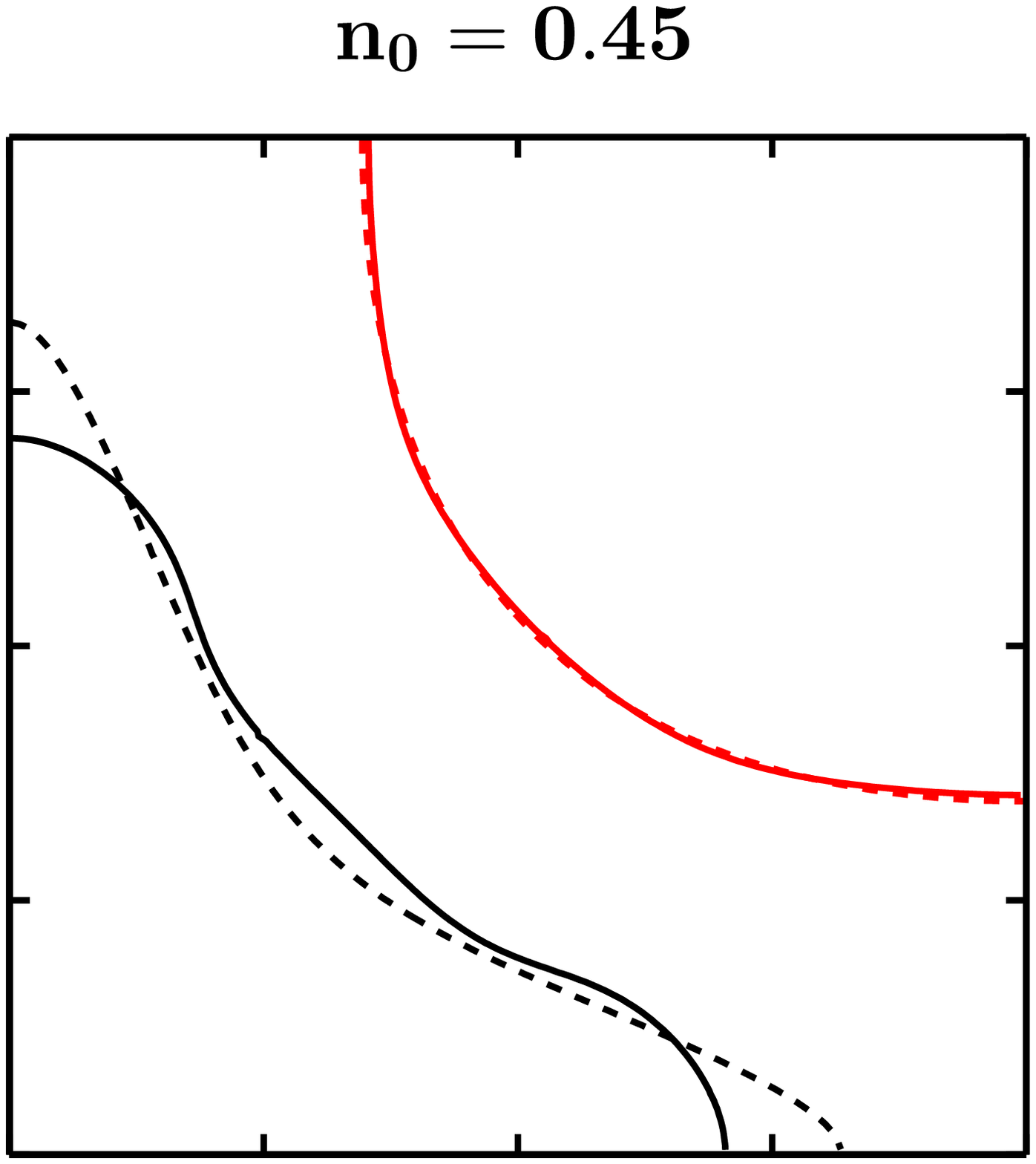}
\\
\includegraphics[scale=0.245]
{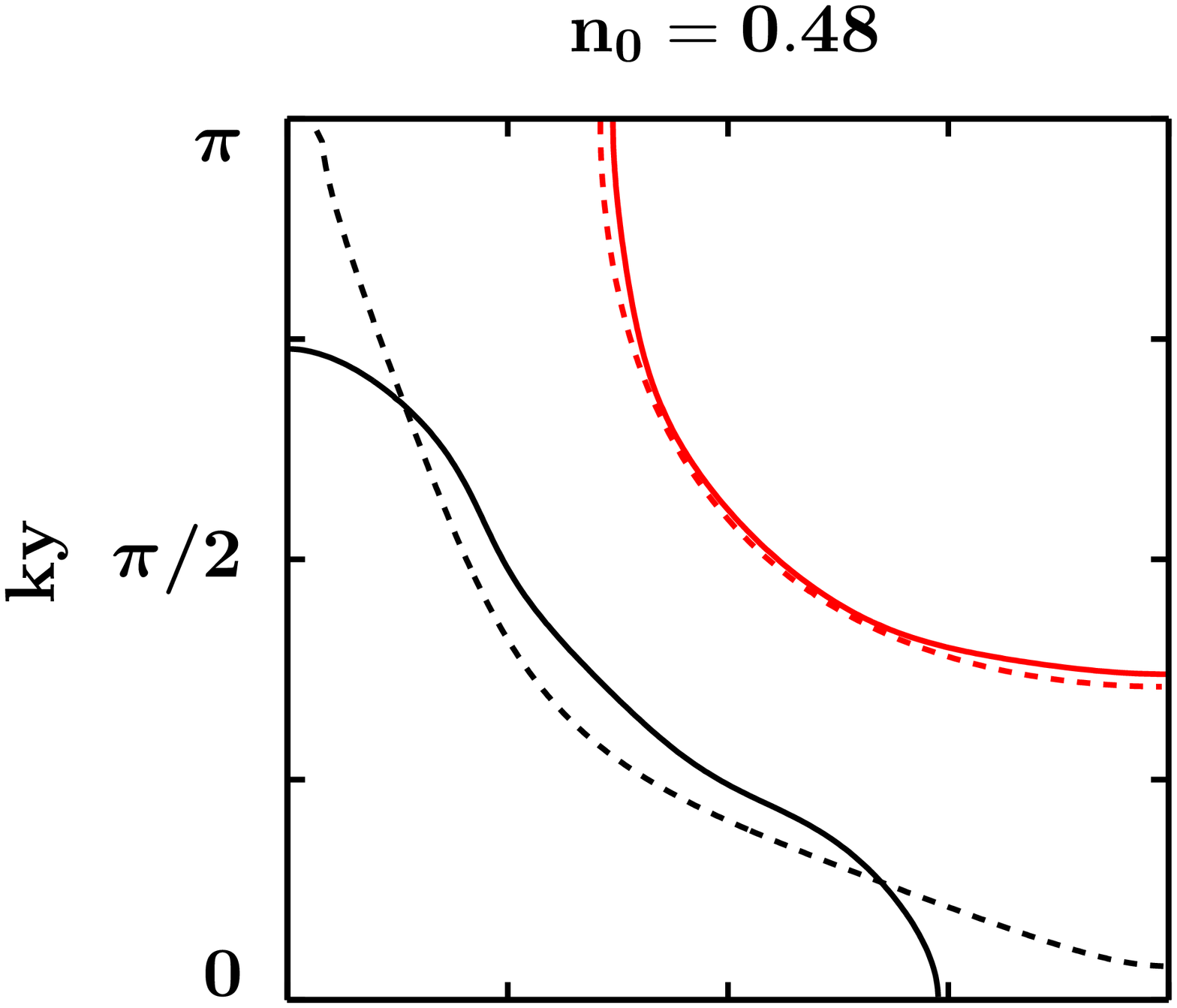}
\vspace{0pt}\includegraphics[scale=0.245]
{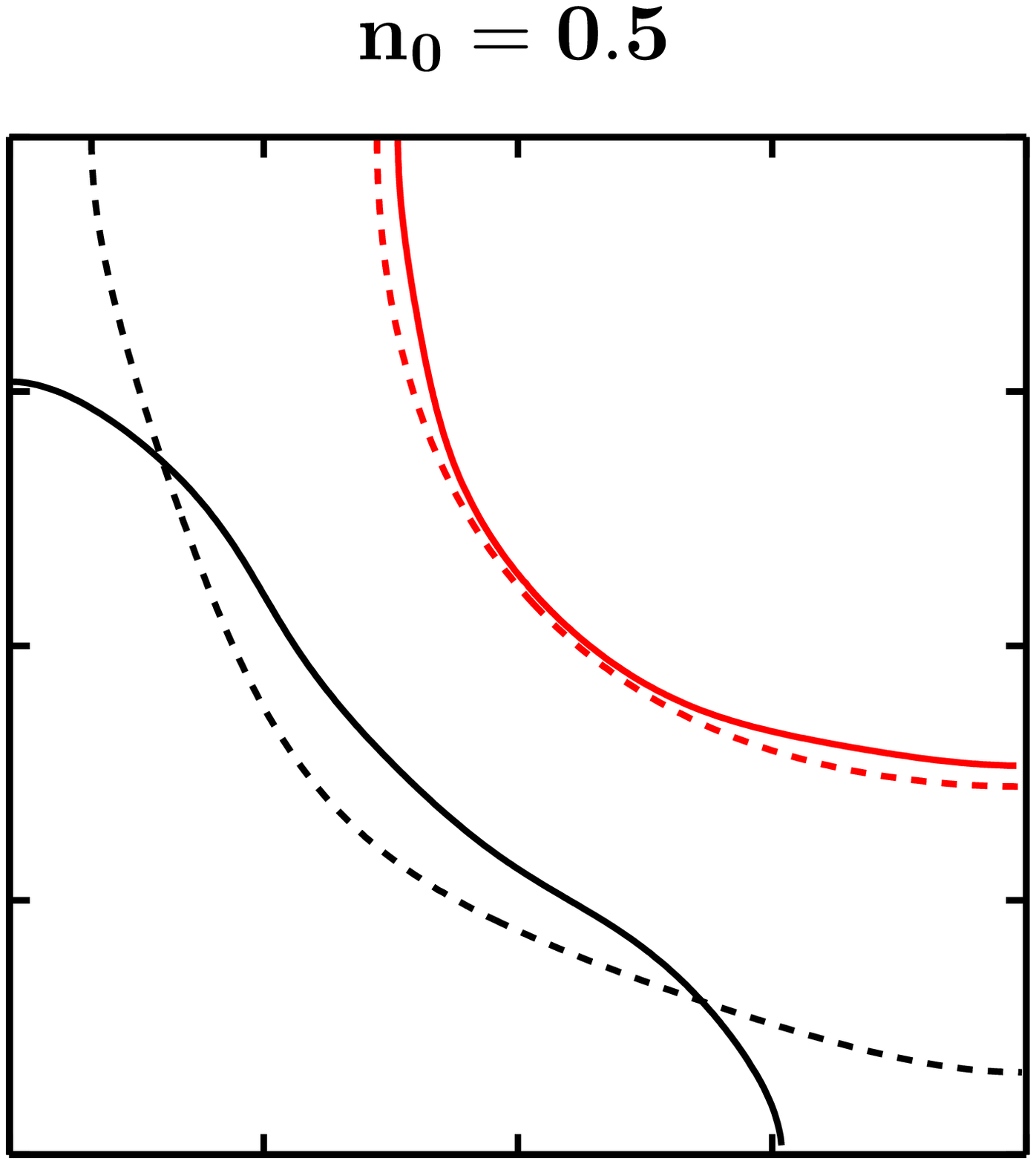}
\\
\vspace{0pt}\includegraphics[scale=0.245]
{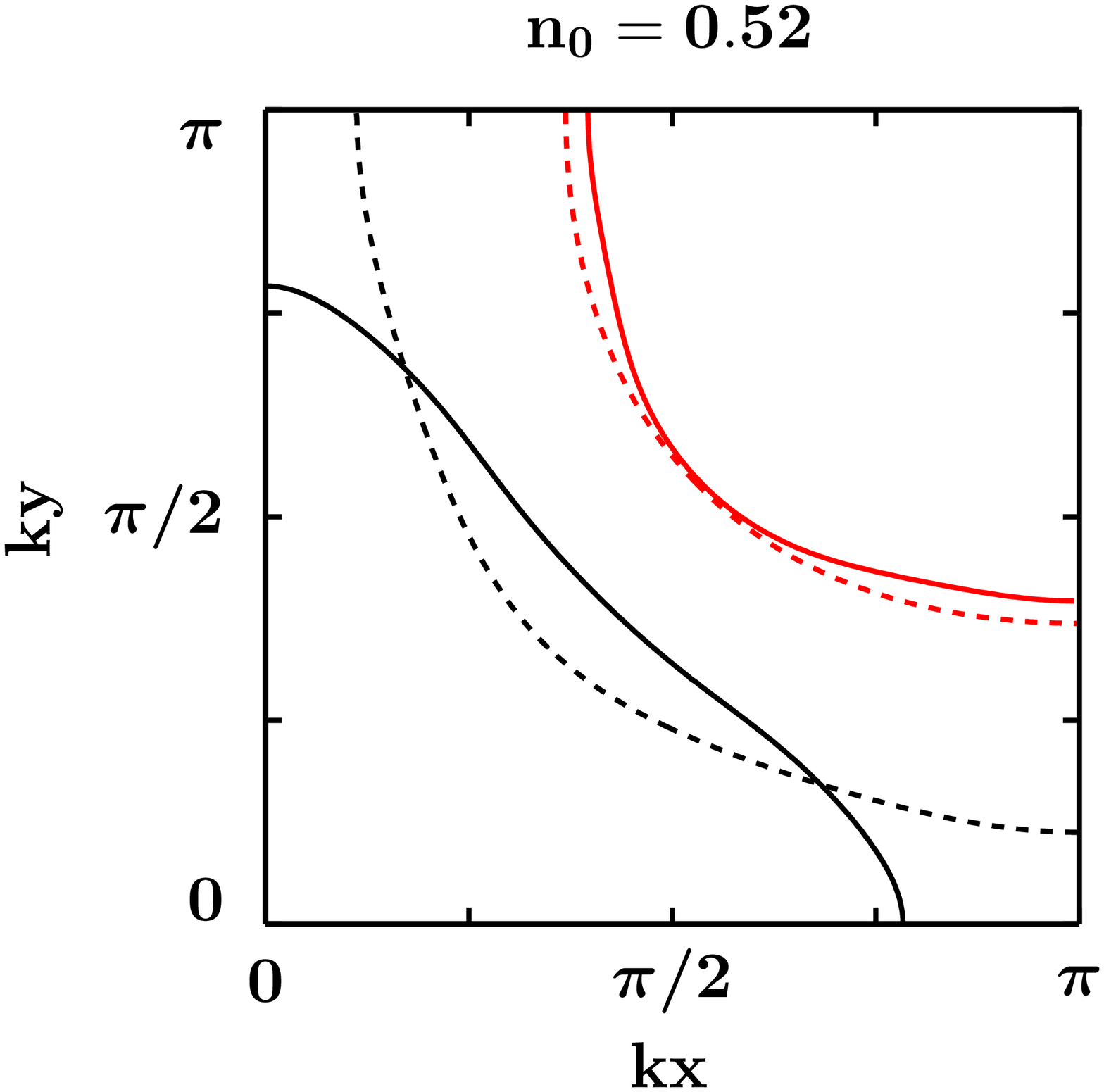}
\includegraphics[scale=0.245]
{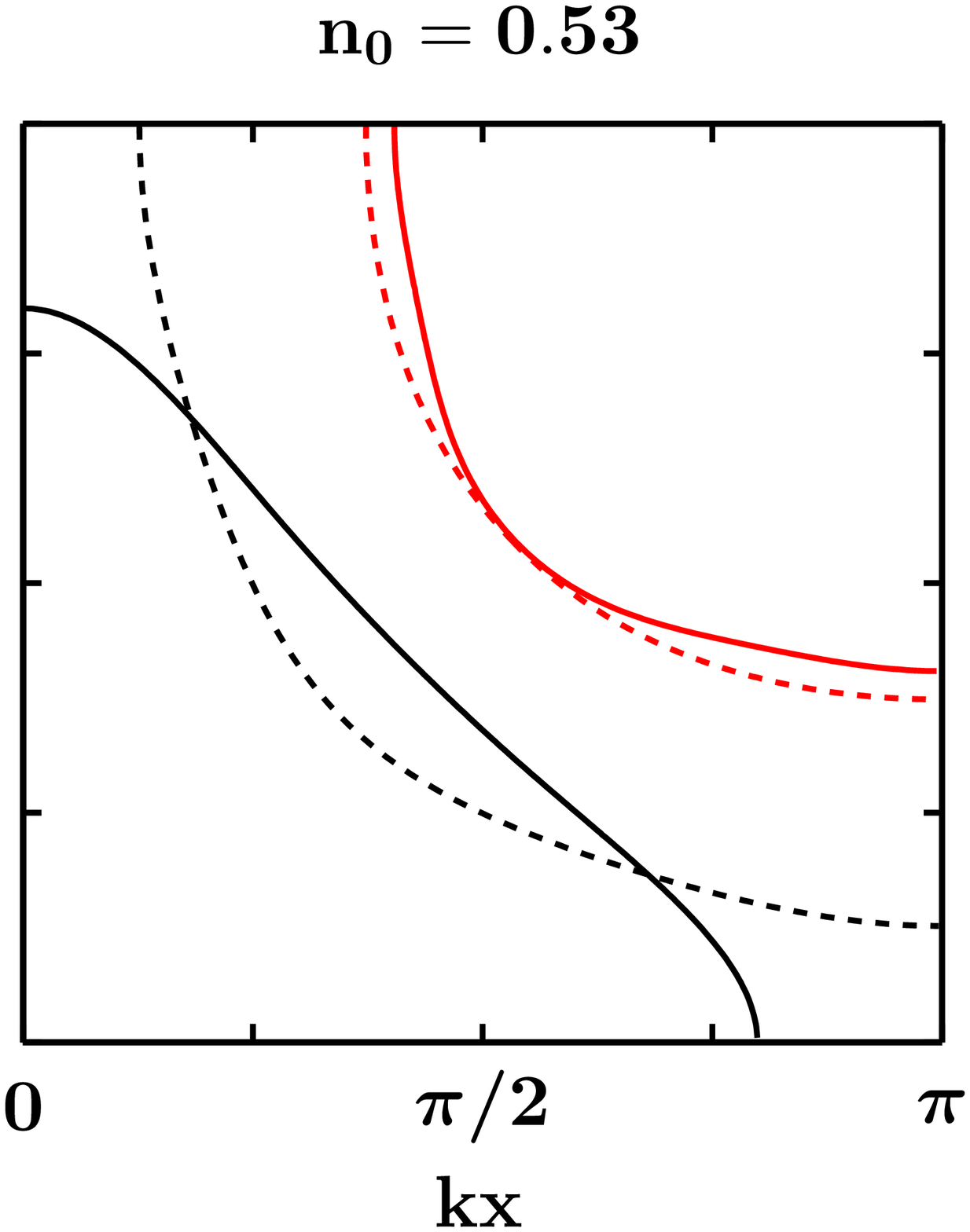}
\caption[Fermi surface deformations.]
{Fermi surface deformations for densities $n_0=0.4$, $0.45$, $0.48$, $0.50$, $0.52$ and $0.53$. 
The local interaction strengths are set to $U=6.0$ and $J=0.8$.
The dashed lines give the initial Fermi edge and the solid lines give the optimized Fermi surface. 
Both bands are occupied in the region between the origin and the solid (dashed) black line. 
In the region between the black and the red lines only the lower band is occupied.
For densities $n_0>0.48$, the optimized inner Fermi edge  has a closed topology while the initial Fermi surface topology is open.}
\label{fig:ps_optimization:fs_deformation_2}
\end{figure}

In Fig.~\ref{fig:ps_optimization:fs_deformation_2}, the Fermi edges of the initial and optimized  SPPS are shown for the densities $n_0=0.4$, $0.45$, $0.48$, $0.50$, $0.52$ and $0.53$. 
The Hubbard/Hund parameters are set to $U=6.0$ and $J=0.8$. 
Although the SPPS $\Psinot$ can have a small but finite magnetization in this parameter regime, we restrict ourselves to a paramagnetic wave function.
The deformation of the inner Fermi surface between $n_k=2$ and $n_k=1$ start for densities $n_0 > 0.4$.
The outer Fermi edge between $n_k=1$ and $n_k=0$ is more robust.
For densities $n_0>0.48$, the optimized inner Fermi edge still has a closed topology while the initial Fermi surface topology is open.
The optimization becomes difficult for densities larger than $n_0= 0.53$.
Alternatively, a particle-hole symmetry can be used to determine the optimal Gutzwiller wave function.\cite{zuMuenster2015}
In this way, we can show that the deviations in the Fermi surface are small for $n_0 \gtrsim 0.6$ where the topology of the optimized Fermi surface becomes open.

\begin{figure}[ht]
 \centering
\vspace{0pt}\includegraphics[width=0.35 \textwidth]
{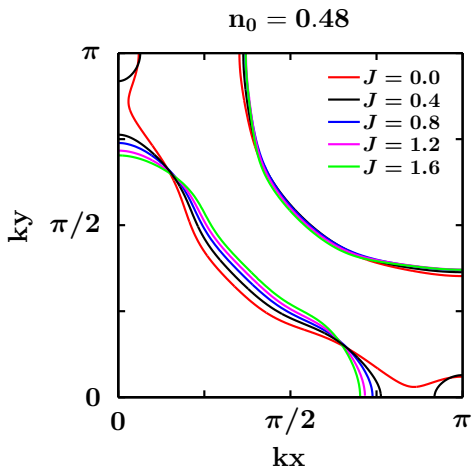}
\caption
{Fermi surface deformations for different interaction strength $J$. 
The density and the interaction strength are set to $n_0=0.48$ and $U=6.0$, respectively.
The deformations increase for larger values of $J$. For $J=0$,   the Fermi surface topology (for $n_k = 2$) is still open.
}
\label{fig:ps_optimization:fs_deformation_j_dependence_2}
\end{figure}

\begin{figure}[t]
 \centering
\vspace{0pt}\includegraphics[scale=0.21]
{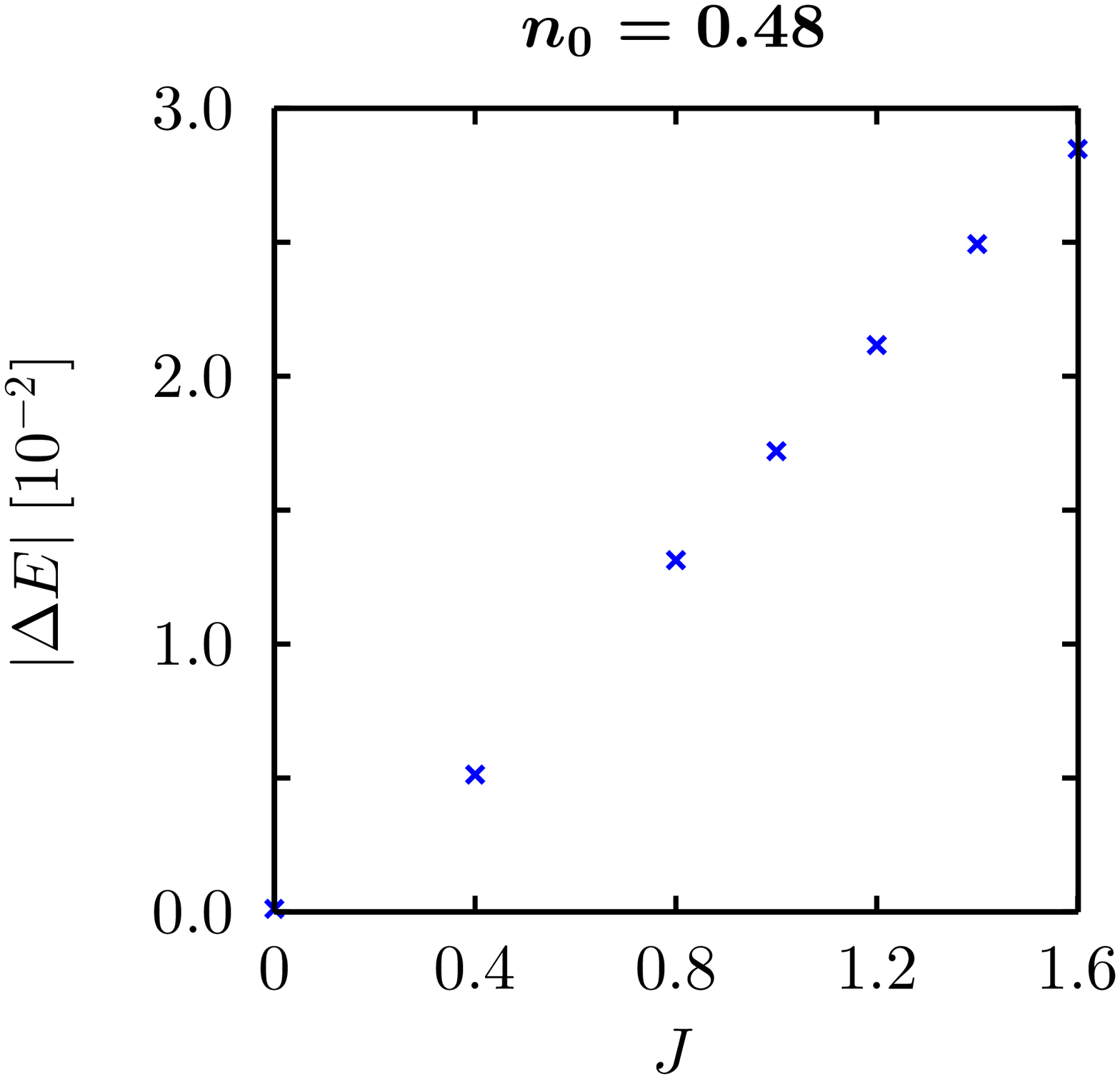}
\vspace{0pt}\includegraphics[scale=0.21]
{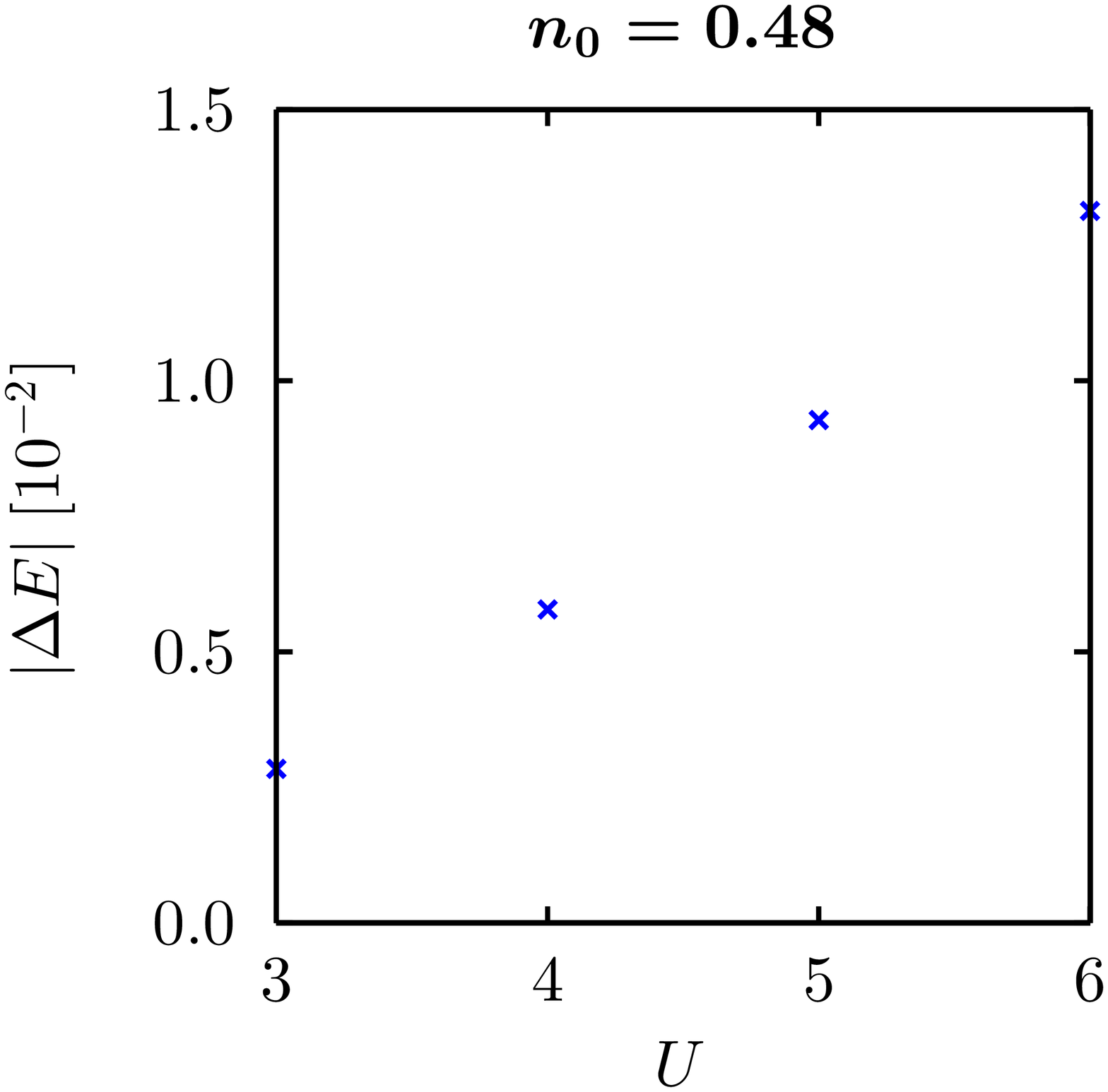}
\caption{Left: The energy gain $\Delta E$ increases linearly in $J$.
Right: Energy gain $\Delta E$ as a function of $U$.}
\label{fig:ps_optimization:fs_deformation_energy_gain}
\end{figure}

The dependence of the Fermi surface deformations on the Hund's-rule coupling $J$ is shown in Fig.~\ref{fig:ps_optimization:fs_deformation_j_dependence_2}.
The Fermi edge for $n_k =2$ remains open for vanishing $J$. 
An increase of the Hund's-rule interaction strength to $J=0.4$ leads to the appearance of small islands in which both bands are filled.
These islands collapse when we further increase the interaction strength to $J=0.8$ so that the Fermi surface becomes closed.
The left panel of Fig.~\ref{fig:ps_optimization:fs_deformation_energy_gain} shows that the energy gain $\Delta E$ increases linearly in $J$ and becomes vanishingly small for $J=0$.
From an energetic point of view, the transition from an open to a closed inner Fermi surface is gradual as a function of $J$.
The existence of intermediate islands also shows that the hopping matrix elements change gradually.

\begin{figure}[ht]
 \centering
\vspace{0pt}\includegraphics[width=0.35 \textwidth]
{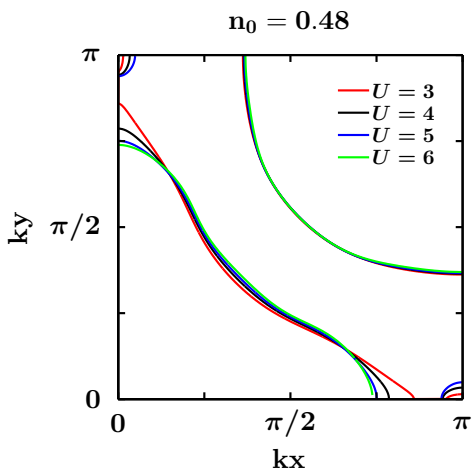}
\caption
{Fermi surface deformation for different interaction strength $U$. 
The density and the Hund's-rule coupling are set to $n_0=0.48$ and $J=0.8$ respectively.
Small islands appear for  $U=3$ in which $n_k = 2$.
}
\label{fig:ps_optimization:fs_deformation_u_dependence_2}
\end{figure}

The change in the Fermi-surface topology as a function of $U$ for $n_0=0.48$ and $J=0.8$ is shown in Fig.~\ref{fig:ps_optimization:fs_deformation_u_dependence_2}.
For an interaction strength $U=3$, the Fermi surface starts to deform from an open to a closed topology and small islands appear.
The islands at the border of the Brillouin zone vanish for $U=6$ again.
The energy gain $\Delta E$ increases linearly in $U$ as shown in the right panel of Fig.~\ref{fig:ps_optimization:fs_deformation_energy_gain}.

In this section, we showed that the interaction-induced Fermi surface deformations are clearly visible and, therefore, do not match the assumptions made in Fermi liquid theory that the Fermi surface of the non-interacting electrons is identical to the quasi-particle Fermi surface.
Moreover, we showed that even the topology of the Fermi surface may change as a function of the Coulomb interaction-strength.
The contributions beyond our second-order approximation still affect the Fermi surface and the density matrix.
However, a higher order expansion of the single-band Hubbard model \cite{Bunemann2012a,jan2015data} showed that the Fermi surface deformations are true features of the Gutzwiller wave function.
Therefore, we can assume that the qualitative findings are valid in all orders of the approximation.

\section{Conclusion and Outlook}
\label{sec:IV}
In this work, we have given a comprehensive derivation of a diagrammatic method that allows us to 
 evaluate multi-band  Gutzwiller wave functions in finite dimensions. Our approach constitutes a 
 systematic improvement of the widely used Gutzwiller approximation which corresponds to 
 the zeroth order of our expansion.  

As our first application, we studied the ferromagnetic phase transition in a two-band Hubbard model on a square lattice 
as a function of the model parameters for various band fillings. 
In general, a large density of states and a strong Hund's-rule exchange favor ferromagnetism. 
In the Gutzwiller wave function, the ferromagnetic order is strongly suppressed so that much larger interaction strength are needed than predicted by the Hartree-Fock solution.
Moreover, the regions in parameter space where non-saturated ferromagnetism occurs are much broader in Gutzwiller theory.
As shown in earlier studies, this gives room for the experimental observations of non-saturated ferromagnetism, e.g., in transition metals such as nickel and iron. It turned out that long-range ferromagnetic order in our model is already  
 well described within the Gutzwiller approximation. 

As a second application, we investigated the inter\-action-in\-duced deformation 
of the Fermi surface.
These effects occur for large interaction strength, when the potential energy of the system is twice as large as the kinetic energy. 
For weaker interactions and small densities, the deformations of the Fermi surface can be neglected.
Close to half band-filling and for special choices of the electron transfer parameters, the interactions can induce a change in the Fermi-surface topology from open to closed constant-energy contours.
These effects are a result of the finite-order diagrams and cannot be seen in the Gutzwiller approximation.

It will be an interesting question for future work whether the Fermi surface deformation can lead to 
symmetry broken phases (Pomeranchuk phase) as in the single-band Hubbard model.
In our two-band model, such a broken rotational spatial symmetry leads to different orbital densities 
which are energetically unfavorable.
Hence, it is an open question if the ground state of our two-band model can have an asymmetric Fermi surface.
 Another open questions concerns the appearance of superconductivity in our model, as seen in earlier work 
 on two-orbital Hubbard models.\cite{PhysRevB.86.014505,1367-2630-15-7-073050,1367-2630-16-3-033001}

\acknowledgments
We like to thank Florian Gebhard for many valuable discussions in all stages of this project. Furthermore we thank Jan Kaczmarczyk for a valuable discussion of his higher-oder study of a single-band Hubbard model.

\appendix
\section{Lattice symmetries}
\label{appendix:lattice_symmetries}

\begin{figure}[ht]
 \centering
\includegraphics[scale=0.30]{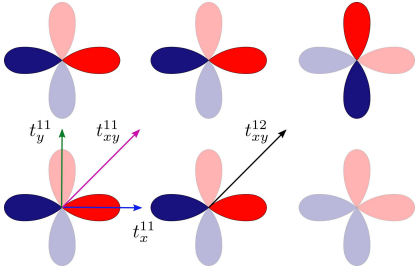}
\caption{Hopping amplitudes}
\label{fig:hamilton:orbitals_hopping1}
\end{figure}

Fig.~\ref{fig:hamilton:orbitals_hopping1} shows the hopping processes to nearest and next-nearest neighbors. 
The amplitude $t_{x}^{11}$ for the transition from the $p_x$ ($\sigma =1$) orbitals on site $i$ to the site $i\pm \rmd x$ equals the amplitude $t_{y}^{22}$ for the transition from the $p_y$ orbitals ($\sigma =2$) on site $i$ to the site $i \pm \rmd y$. 
The same holds for the amplitudes $t_{xy}^{11}$ and $t_{xy}^{22}$ for the transitions between the $p_x$ orbitals on $i$ and $i\pm \rmd y$ and the $p_y$ orbitals on sites $i$ and $i\pm \rmd x$ respectively.
The amplitudes for the hopping processes from $i$ to $i\pm \rmd x\pm \rmd y$ between the $p_x$ orbitals is the same as between the $p_y$ orbitals.
The symmetry of the orbitals does not allow any $p_x$-$p_y$ transition to nearest neighbors. 
For transitions between next-nearest neighbors, the sign of the amplitudes $t_{xy}^{12}$ will change after a rotation of $\pi/2$ so that  $t_{xy}^{12} =-t_{yx}^{12}$.
The $xy$ symmetry of the inter-orbital hopping processes leads to a diagonal local density matrix $\rho_{\sigma,\sigma'}= n^0_{\sigma} \delta_{\sigma,\sigma'}$.
Furthermore, the rotational symmetry of the hopping processes within the same orbitals guarantees that all diagonal entries of the local density matrix are the same, $\rho_{\sigma,\sigma'} = n_0 \Id$.   

The calculation of the on-site Coulomb interaction~(\ref{uuu})
%\begin{eqnarray}
%\label{eq:hamilton:def_on_site_energy}
% \hat{U}_{\rt{int}} = \sum_{\sigma_1,\sigma_2,\sigma_3,\sigma_4} U_{\sigma_1\sigma_2\sigma_3\sigma_4} 
 %\hat{c}_{\sigma_1}^{\dagger} \hat{c}_{\sigma_2}^{\dagger}\hat{c}_{\sigma_3} \hat{c}_{\sigma_4}
%\end{eqnarray}
requires the evaluation of two-particle expectation values of the Coulomb energy 
\begin{eqnarray}
\label{eq:hamilton:coulomb_overlap_integrals0}
 U_{\sigma_1\sigma_2\sigma_3\sigma_4}  = \langle \phi_{\sigma_1} \phi_{\sigma_2} |\hat{V}_{\rt{Coul.}}|\phi_{\sigma_3} \phi_{\sigma_4}\rangle\;,
\end{eqnarray}
with $\sigma_i \in \lbrace p_x, p_y\rbrace$.
These coefficients can be simplified after a decomposition of the $p_x$-$p_y$ orbitals in terms of Laguerre and Legendre polynomials, respectively. 
The explicit derivation of the coefficients can be found in many text books (e.g. in~[\onlinecite{Sugano1970}]) so that we simply state the result.
The matrix representation of the two-particle sector of $\hat{U}_{\rt{int}}$ is given by
\begin{eqnarray}
\label{eq:hamilton:umat}
 U_{{\rm int}} = U \Id + 
\left(
\begin{array}{cccccc}
0 & 0 & 0 & 0 & 0 & J \\
 0 & -3J & 0 & 0 & 0 & 0 \\
 0 & 0 & -2J & -J & 0 & 0 \\
 0 & 0 & -J & -2J & 0 & 0 \\
 0 & 0 & 0 & 0 & -3J & 0 \\
 J & 0 & 0 & 0 & 0 & 0 \\
\end{array}
\right)
\end{eqnarray}
where the standard ordering 
$|\!\uparrow \downarrow , 0\rangle$, 
$ |\!\uparrow ,\uparrow \rangle$, 
$ |\!\downarrow,\uparrow \rangle$, 
$|\!\uparrow , \downarrow \rangle$, 
 $|\!\downarrow ,\downarrow \rangle$, 
$ |0,\uparrow \downarrow \rangle$ 
%\begin{eqnarray}
%|\uparrow \downarrow ,\, 0\rangle, |\uparrow ,\,\uparrow \rangle,\, |\downarrow,\,\uparrow \rangle,\,|\uparrow ,\, \downarrow \rangle,\, |\downarrow ,\,\downarrow \rangle, \,|0,\uparrow \downarrow \rangle
%\end{eqnarray}
 of the two-particle states has been used.  
Thus we can exclude any terms where only one electron switches from a $p_x$ to a $p_y$ orbital or that violates spin-conversation.

The variational coefficients $\lambda_{I_1,I_2}$ in the Gutzwiller correlator 
(\ref{eq:mathematical_definitions:gutzwiller_local})  have the same structure (non-vanishing elements) as the matrix elements of the on-site Coulomb interaction in Eq.~(\ref{eq:hamilton:umat}). 
In the paramagnetic case, the variational coefficients obey a spin-band symmetry.
In the ferromagnetic case, all parameters are symmetric under an interchange of the band index.

\section{HF-operators and diagrammatic weights}
\label{section:appendix:hf_operators}

In this section, we give explicit results for the mapping of an arbitrary operator product to its corresponding HF-operator. 
We use this mapping to define the coefficients $X_{I_1,I_2}$, $Q_{I_1,I_2}$, $Z_{I_1,I_2}$, and $M_{I_1,I_2}$ that give the weights of the internal and external nodes in our diagrammatic expansion.
A detailed derivation of this mappings, its inversion, and a derivation of all coefficients can be 
found in~[\onlinecite{zuMuenster2015}]. 

Consider a product of fermionic creation/annihilation operators $\oa_1 \oa_2 \ldots \oa_n $.
For the corresponding HF-operator $(\oa_1  \oa_2 \ldots \oa_n)^{\HF}$, the evaluation of
\begin{align}
\label{eq:hf_operators:contraction_with_o}
\lbrace \hat{O} (\oa_1  \oa_2 \ldots \oa_n)^{\HF} \rbrace 
\end{align}
shall, by definition, not include contractions between any pairs of operators $\oa_1,\ldots,\oa_n$.
We use the notation
\begin{align}\label{568}
 \lbrace \oa_1 \oa_2 \ldots \oa_n\rbrace_m 
\end{align}
where $m$ denotes the number of internal contractions, e.g., for $m=1$,
\begin{align}
 \lbrace \oa_1  \oa_2 \ldots \oa_n\rbrace_1 &= \sum_{j<k} (-1)^{j+k+1} \lbrace \oa_j \oa_k \rbrace \\ \nonumber
 & \quad \oa_1 \ldots\oa_{j-1}\oa_{j+1} \ldots \ldots\oa_{k-1}\oa_{k+1}\ldots\oa_n .
\end{align} 
Each internal contraction reduces the number of operators by two.
With the abbreviation (\ref{568}) we can give the following closed expression for the HF-operator 
of an arbitrary operator
\begin{align}
\label{eq:hf_operators:redefine_hf_3}
 \left(\oa_1  \oa_2 \ldots \oa_n \right)^{\HF} &= \sum_{k=0}^{[n]}  (-1)^k 
\lbrace a_1  a_2 \ldots a_n \rbrace_k \;,
\end{align}
where $[n]$ denotes the next smallest even number.
This result agrees with the expressions for the definition of the HF-operators in~[\onlinecite{Bunemann2009}].

In our diagrammatic expansion, we must express all operators  $\Cd_{K_1}\C_{K_2}$ in terms of HF-operators,
\begin{align}
 \label{eq:hf_operators:hf_operator_expansion_0}
 \Cd_{K_1} \C_{K_2} &= 
\sum_{I_1,I_2} X_{I_1,I_2}^{K_1,K_2} \left( \Cd_{I_1} \C_{I_2} 
\right)^{\rt{HF}} \;.
\end{align}
Let us contract both sides with an arbitrary operator $\hat{O}$. 
In order to determine the value of the coefficient $X^{K_1 K_2}_{I_1,I_2}$, we have to evaluate the term where the operators $\Cd_{I_1}$ and $\C_{I_2}$ with $I_1 \subset K_1$, $I_2 \subset K_2$ form the external contractions with $\hat{O}$.
We need to shift the operators that are reserved for the external contractions to the front of the operator $\hat{O}$, and contract all remaining operators internally. 
The operator $\hat{O}$ can be chosen without any restrictions and just indicates which operators are reserved for an external contraction. 
The contraction with $\hat{O}$ can be carried out symbolically by a replacement of $\Cd_{I_1} \C_{I_2}$ with the HF-operators $( \Cd_{I_1} \C_{I_2} )^{\rt{HF}}$. 
Thus, we get
\begin{align}
\label{eq:hf_operators:hf_operator_expansion_x12}
 X^{K_1,K_2}_{I_1,I_2}  &= 
 \sigcl{I_1}{K_1} \sigar{K_2}{I_2}
\lbrace \Cd_{K_1 \setminus I_1 } \C_{K_2 \setminus I_2} \rbrace \;,
\end{align}
where we introduced the following symbols to indicate that we consider a sign change after a reordering of creation or annihilation operators.
\begin{align}
 \sigcr{I}{J} &\quad \rt{sign after splitting } \Cd_{I \cup J} \to \Cd_{I} \Cd_{J} \;,\\ \nonumber
    \sigar{I}{J} &\quad \rt{sign after splitting } \C_{I \cup J} \to \C_{I} \C_{J} \;,\\ \nonumber
 \sigcl{J}{I} &\quad \rt{sign after splitting } \Cd_{I \cup J} \to \Cd_{J} \Cd_{I} \;,\\ \nonumber
 \sigal{J}{I} &\quad \rt{sign after splitting } \C_{I \cup J} \to \C_{J} \C_{I} \;,\\ \nonumber
   \sigcu{I}{J} &\quad \rt{sign after merging } \Cd_{I} \Cd_{J} \to \Cd_{I \cup J} \;,\\ \nonumber
  \sigau{I}{J} &\quad \rt{sign after merging } \C_{I} \C_{J} \to \C_{I \cup J} \;.
\end{align} 
All operators are assumed to be normally ordered before and after the process. 
The reversed ordering of the annihilation and creation operators ensures that
\begin{align}
 \sigcr{I}{J} =  \sigal{J}{I} = \sigcu{I \setminus J}{J} = \sigau{J}{I 
\setminus J}\;.
\end{align}

As a next step we transform the square of the local Gutzwiller operator into a sum of HF-operators,
\begin{align}
\label{eq:hf_operators:hf_operator_expansion_gw_1}
 P^{\dagger}P &=  
 \sum_{I_1, 
I_2} X_{I_1,I_2} \left( \Cd_{I_1} \C_{I_2} \right)^{\rt{HF}}\;.
\end{align}
An application of the mapping~(\ref{eq:hf_operators:hf_operator_expansion_x12}) gives
\begin{align}
\label{eq:definition:X}
 X_{I_1, I_2} 
  &= \!\!\!\!\!\!  \sum_{\substack{J_1\subset \bar{I}_1, J_2 \subset \bar{I}_2 \\ |J_1|=|J_2| } } \!\!\!\!\!\!
  \lbrace \Cd_{J_1} \C_{J_2 } \rbrace \sigcu{I_1}{J_1} \;\sigau{J_2}{I_2 } 
  \\ \nonumber &\phantom{=} \!\!\!\!\!\!\!
\sum_{\substack{J_3 \subset \\ (I_1 \cup J_1) \cap (I_2 \cup J_2)}}   \!\!\!\!\!\!\!\!\!\! (-1)^{|J_3|}
\lbarp{(I_1\cup J_1) \setminus J_3 }{(I_2 \cup J_2) \setminus J_3}   \;
\\ \nonumber &\phantom{=}\;\;\;\;\;\;\;\;
\sigcr{I_1\cup J_1}{J_3}\; \sigal{J_3}{I_2 \cup J_2}
\;.
\end{align}

The external nodes defined in Eq.~(\ref{eq:definition:q_operator}) can be written as 
\begin{align}
\hat{Q}(\hat{O}) =  \sum_{I_1,I_2} K_{I_1,I_2}(\hat{O})  | I_1 \rangle \langle I_2 |
\end{align}
with
\begin{align}
  K_{I_1,I_2}(\hat{O}) =  \sum_{I_3,I_4} \lambda_{I_3,I_1} 
\lambda_{I_4,I_2} \langle I_3 | \hat{O} | I_4 \rangle \;.
\end{align}
Then, we can use the HF mapping to find
\begin{align}
 \hat{Q}(\hat{O}) = \sum_{I_1,I_2} Q_{I_1,I_2}(\hat{O}) \left( \Cd_{I_1} \C_{I_2} 
\right)^{\rt{HF}}\;,
\end{align}
which still depends on the operator $\hat{O}$.
Here,
\begin{align}
 Q_{I_1,I_2}(\hat{O}) &= \!\!\!\!\!\!
 \sum_{\substack{J_1 \subset \bar{I}_1, J_2 \subset I_2 \\ |J_1|=|J_2| } } \!\!\!\!\!\!
\lbrace \Cd_{J_1} \C_{J_2} \rbrace
\sigcu{I_1}{J_1} 
\sigau{J_2}{I_2}
\\ \nonumber &\phantom{=} \!\!\!\!\!\!\!\!\! 
 \sum_{\substack{ J_3 \subset \\  (I_1\cup J_1)\cap(I_2\cup J_2)}} \!\!\!\!\!\!\!\!\!\! (-1)^{|J_3|}  
 K_{(I_1\cup J_1)\setminus J_3}^{(I_2\cup J_2)\setminus J_3}(\hat{O}) 
\\ \nonumber &\phantom{=} \;\;\;\;\;\;\;\;
\sigcr{I_1\cup J_1}{J_3}  \sigal{J_3}{I_2\cup J_2}  \;,
\end{align}
where the coefficients $K_{I_1,I_2}$ play the role of the coefficients $\bar{\lambda}_{I_1,I_2}$ in Eq.~(\ref{eq:definition:X}).

The internal nodes defined in Eq.~(\ref{eq:lct:exponential_form_internal}) are given by
\begin{align}
  \tilde{G} = \!\!\!\!\!\!  \sum_{ \substack{I_1,I_2 \\ |I_1|,|I_2| >0}} \!\!\!\!\!\!   Z_{I_1,I_2} \tilde{C}^{\dagger}_{I_1} \tilde{C}^{\phantom{\dagger}}_{I_2} \;,
\end{align}
with
\begin{align}
 Z_{I_1,I_2} = \sum_{m>0} \frac{(-1)^{m+1}}{m} \!\! \sum_{\substack{\{(J_1^s,J_2^s)\} \\ s=1,\ldots,m}} \!\!\! 
\Sigma[J_1^s,J_2^s] \prod_{s=1}^m X_{J_1^s,J_2^s} \;,
\label{eq:B16}
\end{align}
where the sum in eq.~(\ref{eq:B16})
runs over all (disjunct) partitions $\{(J_1^s,J_2^s)\}$ of the set $(I_1,I_2)$ such that
\begin{align}
 \bigcup_s J_1^s = I_1 \rt{ and }  \bigcup_s J_2^s = I_2 \;,
\end{align}
and $\Sigma[J_1^s,J_2^s]$ gives the sign which is necessary to convert the operator product $\prod_{s} \Cd_{J_1^s} \C_{J_2^s}$ into normal order again.
The weight of the external nodes in Eq.~(\ref{eq:lct:exponential_form_internal2}) can be written as
\begin{align}
  \tilde{M} = \sum_{I_1,I_2 } M_{I_1,I_2} \tilde{C}^{\dagger}_{I_1} \tilde{C}^{\phantom{\dagger}}_{I_2} \;,
\end{align}
with
\begin{align}
   M_{I_1,I_2} = Q_{I_1,I_2}-\sum_{m=1} (-1)^{m+1} \!\!
\sum_{\substack{\{(J_1^s,J_2^s)\} \\ s=0,\ldots,m}} \!\!  \Sigma[J_1^s,J_2^s] 
\\ \nonumber
\quad
 Q_{J_1^0,J_2^0}  \prod_{s=1}^{m} X_{J_1^s,J_2^s}  \;.
\end{align}

\section{Previous approaches}
\label{section:appendix:previous_approaches}
The diagrammatic analysis of the single-band case which includes the HF-operators was first worked 
out in~[\onlinecite{Gebhard1990,Gebhard1990}].
There, the correlator  $\tilde{A} = x_d \tilde{n}_{\uparrow}\tilde{n}_{\downarrow}$ is employed with $x_d$ being the only non-vanishing $X_{I_1,I_2}$ coefficient. 
In this case, we can set 
\begin{eqnarray}
 1+x_d  \tilde{C}^{\dagger}_{\uparrow, \downarrow} \tilde{C}_{\uparrow, 
\downarrow} = \exp(x_d \tilde{C}^{\dagger}_{\uparrow, \downarrow} 
\tilde{C}_{\uparrow, 
\downarrow})\;.
\end{eqnarray}
When we work with multiple bands or if we allow terms like $x_{\uparrow} \tilde{c}^{\dagger}_{\uparrow}\tilde{c}^{\phantom{\dagger}}_{\uparrow} $ in our correlator, we need to re-exponentiate our Gra\ss mann operators. 
This has been overlooked in the deviation of the LCT in the multi-band case in~[\onlinecite{Bunemann2009}] although the problem was already noticed in~[\onlinecite{Buenemann1997}]. 

In principle, the transformation of the ladder operators to HF-operators is not a necessary step for the transformation to Gra\ss man variables.
After all operators have been brought to normal ordering every operator in the numerator (denominator) will appear only once. 
The operators can be mapped to Gra\ss mann variables 
$\hat{c}_{i\sigma}^{(\dagger)} \to \tilde{c}_{i\sigma}^{(\dagger)}$
%\begin{eqnarray}
% \hat{c}_{i\sigma}^{\dagger} \to \tilde{c}_{i\sigma}^{\dagger}\;, \quad& \hat{c}_{i\sigma} \to \tilde{c}_{i\sigma} \
% \;,
%\end{eqnarray}
with vanishing anticommutator $\lbrace \tilde{c}_{i\sigma}^{\dagger},\tilde{c}_{i\sigma}\rbrace = 0$.
In contrast to the Gra\ss mann variables defined in Eq.~(\ref{eq:lct:grassmann}), the local contractions are still finite.
The external and internal nodes can be defined as in Eq.~(\ref{eq:nodes_intermedeate}).

In this approach, the diagrams will also include local lines.
That means that we need to include the summation of an arbitrary number of (directly connected) nodes sitting on the same site, in order to sum up all local contractions.
For the single-band model the diagrammatic expansion of this case is derived 
in~[\onlinecite{Metzner1987,Metzner1988,Metzner1989a,Kollar2001}],
where the cases of one and infinite dimensions are treated analytically.  
A similar approach for a three-flavor system with an Gutzwiller correlator of the form  $1+\alpha \hat{n}_1 \hat{n}_2 \hat{n}_3$ can be found in~[\onlinecite{Rapp2008}], where the local contractions are still present. 
The transformation to an exponential function is again trivial.

% \bibliography{bhm_paper_added.bib}
\bibliography{References/library.bib}

\end{document}